\documentclass[twocolumn]{aastex631}

\usepackage{natbib}
\usepackage{amsmath, mathtools, amsthm, amssymb}
\usepackage{hyperref}
\usepackage{graphicx}
\usepackage{threeparttable}

\shorttitle{Double-exponential disk galaxies}
\shortauthors{Sarkar et al.}

\begin{document}

\title{Finding massive double-exponential disk galaxies with extended low surface brightness stellar disk - an IllustrisTNG exploration}

\email{suchira.sarkar@iucaa.in}
\email{kanak@iucaa.in}

\author{Suchira Sarkar}
\affil{Inter-University Centre for Astronomy \& Astrophysics, Pune-411007, Maharashtra, India}

\author{Kanak Saha}
\affil{Inter-University Centre for Astronomy \& Astrophysics, Pune-411007, Maharashtra, India}

\begin{abstract}
We study massive disk galaxies (stellar mass $>=10^{11}\mathrm{M_{\odot}}$) at z=0 from IllustrisTNG simulation to detect galaxies that contain two exponential stellar disks - a central high surface brightness (HSB) disk surrounded by an extended low surface brightness (LSB) envelope. This is motivated by the observation of several giant LSB galaxies (Malin 1, UGC 1378, UGC 1382 etc), reported in the literature, showing such complex morphology. Studying such systems can help us to understand the mass assembly process and growth of stellar disks in massive galaxies and thus can shed light on massive galaxy formation and evolution scenarios. We use the high-resolution IllustrisTNG50 data and perform S\'{e}rsic plus exponential profile modeling on the idealized, synthetic SDSS g, r-band images of the massive disk galaxies using GALFIT.  We identify 7 disk galaxies (12\% of the parent sample of disk galaxies) that are best represented by a central S\'{e}rsic plus a central HSB disk surrounded by an extended LSB disk. The radial scale lengths of the LSB disk lie in the range of $\sim$ 9.7-31.7 kpc, in agreement with that found in the literature. We study the star formation properties of these simulated double-disk galaxies to understand the distribution of these from blue star-forming to red-quenched region. Some of these double-disk galaxies display a characteristic minima in their (g-r) color radial profiles. The double-disk galaxies are found to lie within $\sim$ 1.5-$\mathrm{\sigma}$ region of the Baryonic Tully-Fisher relation from observation.
\end{abstract}

\keywords{Galaxies (573) --- Galaxy disks(589)--- Giant galaxies(652) --- Low surface brightness galaxies(940) ---Galaxy formation(595)}

\section{Introduction}

The formation and evolution of the most massive disk galaxies (total stellar mass $\sim$$\mathrm{10^{11}M_{\odot}}$) in the nearby universe (z $<$ 0.1) represents an important open question in the current research in extra-galactic astrophysics. Studying the structural analysis of such galaxies is important since it encodes the formation and evolution of them across redshift. The radial surface brightness profile of the stellar disk in galaxies is typically observed to be exponential in nature, characterized by a central surface brightness value and a radial scale length \citep{deVaucouleurs1959,Freeman1970,Fall1980}. Although the origin of the exponential profile has widely been studied in literature, it still remains an intriguing question till date. Very interestingly, many disk galaxies have a complex morphology- they are best described by two exponential functions of different radial scale lengths instead of a single exponential function. The outer exponential could be steeper as well as shallower than the inner exponential profile. Several possible formation mechanisms, including external as well as secular evolution, have been discussed in this context. \cite{HunterElmegreen2006} studied UBV imaging of a large sample of irregular galaxies and attributed the presence of a double-exponential profile to the difference in the rate of star formation in the inner \& outer disk. Using N-body simulation, \citet{Penarrubiaetal2006} showed that the tidal disruption of gas-rich dwarf satellite on a prograde orbit, coplanar with the host galaxy disk, can settle into an outer, extended exponential disk around the host. Simulations also show that a piece-wise exponential could form in spirals by minor mergers \citep{,Youngeretal2007,Erwinetal2008} and even in S0 galaxies through major mergers \citep{Borlaffetal2014}. Scattering of stars by spiral arms \citep{SellwoodBinney2002}, biased stochastic scattering of stars in disk \citep{ElmegreenStruck2016}, presence of a star formation threshold along with stellar radial migration \citep{Roskaretal2008} are some of the other mechanisms discussed in the literature. Thus, the surface brightness profile of a disk galaxy encodes the details of the mechanisms that drive disk structural evolution across cosmic time and provide constraints on the galaxy formation concepts.

Several giant low surface brightness galaxies (GLSBs) in the nearby universe, a class of very massive disk galaxies (total stellar mass $\sim 10^{11} \mathrm{M_{\odot}}$), have also been shown to be best described by two stellar disk profiles. The stellar disks are found to be best modeled by a central high surface brightness (HSB) core surrounded by an outer, extended low surface brightness (LSB) envelope. A stellar disk is termed a low surface brightness disk if its central surface brightness is typically $>$22.5 $\mathrm{mag \ arcsec^{-2}}$ in the optical B-band \citep{McGaugh1996, Rosenbaumetal2009}. Giant LSBs harbor an extended LSB stellar disk of exponential scale length more than about 10 kpc \citep{Sprayberry1995, Das2013, Saburovaetal2023, Zhuetal2023}. Malin 1, a massive ($\mathrm{log(M_{\star}/M_{\odot})}$=11.9 \citep{Sahaetal2021}), prototypical giant LSB galaxy was first observed to contain a large stellar disk extending up to $\sim$ 100 kpc in radius \citep{MooreParker2006, Galazetal2015}, with an extrapolated central surface brightness of $\sim$ 25.5 mag arcsec$^{-2}$ in V-band \citep{Bothunetal1987, ImpeyBothun1989}. Later, \citet{Barth2007} discussed that the morphology of Malin 1 should be best described using a central bulge, bar, a HSB disk surrounded by an outer extended LSB disk, which is photometrically decoupled from the inner disk. \citet{Sahaetal2021} also studied the structural decomposition of Malin 1 using HST and CFHT (Canada-France Hawaii Telescope) g-band observation and found an extended LSB structure containing spiral arms. 
\citet{Boissieretal2016} modeled Malin 1 in u,g,i,z bands using a S\'{e}rsic plus exponential disk extending up to around 130 kpc. They found the angular momentum of the extended disk of Malin 1 to be almost 20 times that of the Milky Way, making it an extremely unique galaxy. Very recently, \citet{Johnstonetal2024} studied the nuclear region of Malin 1 using observation from VLT/MUSE, and claimed that the center of Malin 1 has a double nucleus, under the form of two central star clusters. 
\citet{Galazetal2024} observed molecular CO gas in Malin 1 for the first time using ALMA observation, supporting \citet{Junaisetal2024} observations that the LSB disk is forming stars at several places. 
However, a detailed understanding of what formed the hybrid morphology and the extended LSB disk remains elusive to date. 
Several other GLSBs with a double-exponential disk morphology, although less extreme than Malin 1, were studied over the last decade, e.g, UGC 1382 \citep{Hagenetal2016}, NGC 2841 \citep{Zhangetal2018}, UGC 1378 \citep{Saburovaetal2019} etc. 

Interestingly, \citet{Pandeyetal2022} reported a lenticular galaxy IZw 81 in the Bootes void region, relatively smaller in radial extent, containing a HSB plus LSB morphology. The detailed structural parameters of all the galaxies mentioned above are given in Section 4. 
 
Usually, the most massive galaxies are believed to undergo a morphological transformation from disk to ellipticals across cosmic time, and thus, the morphology of such galaxies in the nearby universe should be dominated by ellipticals. Nevertheless, a significant number of massive disk galaxies are observed in the nearby universe, including the class of GLSBs \citep{Jacksonetal2020, Jacksonetal2022}. 
In this scenario, the physical process by which such galaxies build their photometrically decoupled outer extended, LSB stellar disk, as well as the survival of such disks till z=0, pose intriguing questions. The outer disk in these GLSBs could be built by accretion of stars from dwarf satellites, by stars migrated from the inner galaxy via secular evolution process, or, there could be in-situ star formation in the LSB disk itself (\citet{Hagenetal2016}; \citet{Zhangetal2018}). Therefore, studying such systems is important. This will eventually shed light on the growth of stellar disk and the mass assembly process of the massive disk galaxies across cosmic time. 
But the lack of observational detection of a statistically significant sample of such galaxies prevents us from a generic understanding of their formation mechanism. Such systems could be genuinely rare to form, or, this could be due to the limitation of the detection of the faint outskirt. 

The formation of LSB galaxies containing single exponential stellar disk has been studied in detail in cosmological hydrodynamical simulation suits, e.g., using NIHAO \citep{DiCintioetal2019}, EAGLE \citep{Kulieretal2020}, IllustrisTNG \citep{PerezMontanoetal2022} etc. \citet{Zhuetal2018, Zhuetal2023} studied GLSBs with single exponential stellar disk using IllustrisTNG simulation. However, a systematic study of massive double-exponential disk galaxies with outer, extended LSB disk remained little explored.
The advent of deep and wide photometric data from the ongoing observations such as using the Subaru telescope (Hyper Suprime Cam Subaru Strategic Program \citep{Aihara2022}), Dragonfly telescope \citep{Abrahametal2014} are supposed to increase the detection limit of the low surface brightness regime of galaxies unprecedentedly. The upcoming survey by LSST (Large Synoptic Survey Telescope \citep{Ivezicetal2019}) is supposed to give data for billions of galaxies as well. Therefore, we could expect this to enable the detection of a statistically significant sample of double-disk galaxies with an outer, extended LSB disk. This will help us to understand how common such galaxies are in the local universe and constrain their formation scenario as well. In this context, a theoretical effort to study massive double-exponential disk galaxies from large volume cosmological simulations is very timely.

In this paper, we use publicly available data from the IllustrisTNG simulation suit, a cosmological gravo-magneto-hydrodynamical simulation \citep{Pillepichetal2018a,Pillepichetal2018b}. We make use of its highest resolution run, i.e., TNG50-1 \citep{Nelsonetal2019, Pillepichetal2019}, and perform morphological bulge+disk decomposition to find such galaxies.
We note that \citet{Zhuetal2023} primarily did a bulge plus an exponential disk modeling (S\'{e}rsic plus single exponential) to determine GLSB galaxies and then also explored if just a two-disk fitting is better than a S\'{e}rsic plus disk fitting. They found that both models are exactly equally good for representing their sample of GLSBs and, therefore, chose to describe the galaxies using bulge and single disk. Here, we aim to determine galaxies that have a complex, layered morphology- a bulge/bar, an inner HSB disk, and an outer, extended LSB disk, which has not been theoretically explored yet.
The paper is organized as follows. We introduce the simulation details in Section 2, the method to detect the double-disk galaxies from the simulation data in Section 3, the structural and other physical properties of the double-disk galaxies in Section 4, and summarize our results in Section 5.

\section{Data}
\begin{figure}
\centering
\includegraphics[height=2.3in,width=3.2in]{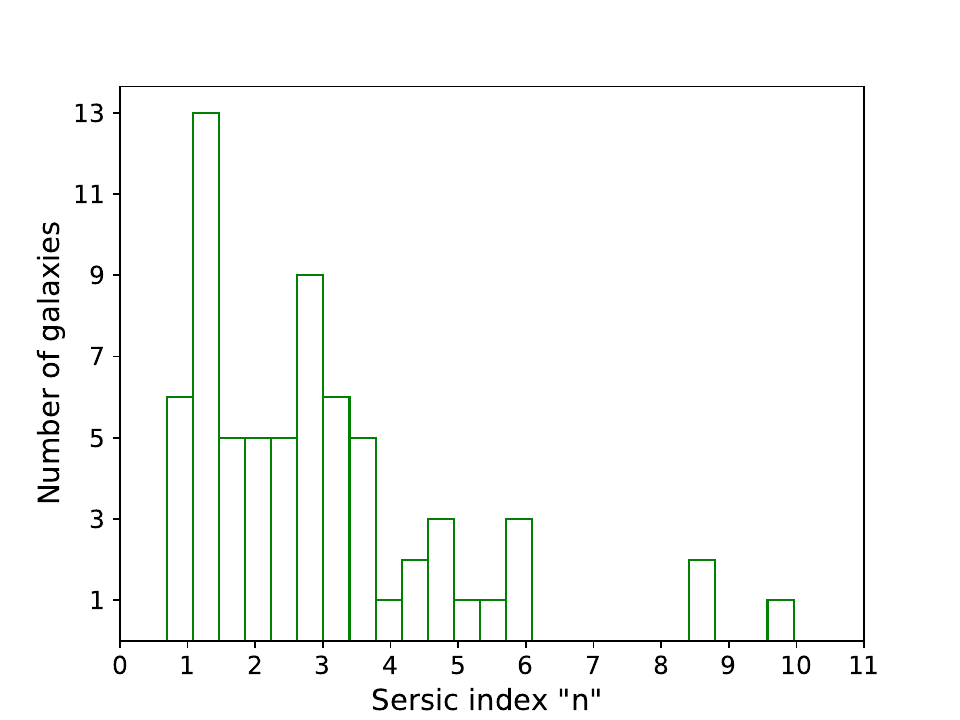}
\caption{Distribution of the S\'{e}rsic index (n) of the sample of elliptical and disk galaxies, obtained from single S\'{e}rsic modeling by GALFIT. The no. of bins used to show the distribution is 24. We used $n<2.5$ criteria to select disk galaxies. We note that several visually identified spiral galaxies spuriously give $n>2.5$ here. }

\label{fig:fig1}
\end{figure}

We use publicly available data from the IllustrisTNG simulation suit \citep{Marinaccietal2018, Naimanetal2018, Nelsonetal2018, Pillepichetal2018a, Springeletal2018}. It is a suite of cosmological gravo-magneto-hydrodynamical simulations run using the moving-mesh code AREPO \citep{Springel2010}. It is the successor of the original Illustris simulation \citep{Geneletal2014, Vogelsbergeretal2014a, Vogelsbergeretal2014b, Sijackietal2015}. The galaxy formation model employed in the TNG simulation is outlined in \citet{Weinbergeretal2017} and \citet{Pillepichetal2018a}, and is calibrated to reproduce the z=0 galaxy stellar mass function, galaxy stellar-to-halo mass relation, galaxy stellar mass-size relation etc. The TNG galaxy formation model is an updated and improved version of the model used in the Illustris project. The IllustrisTNG simulations are run using the cosmologically motivated initial conditions, assuming an updated cosmology consistent with the Planck results \citep{Planck2016}, where $\Omega_{\Lambda,0}$=0.6911, $\Omega_{m,0}$=0.3089, $\Omega_{b,0}$=0.0486, h=0.6774, $\sigma_{8}$=0.8159, $n_{s}$=0.9667. The IllustrisTNG project consists of three simulation volumes. The physical simulation boxes have cubic volumes of roughly 50, 100, and 300 Mpc side lengths, and the simulations are referred to as TNG50, TNG100, and TNG300, respectively. Here we use the data from the highest resolution run, i.e., TNG50-1 \citep{Nelsonetal2019, Pillepichetal2019} which realizes a simulation cube with a volume of 51.7$^{3}$ Mpc$^{3}$, and mean baryonic mass and dark matter particle mass resolution of 8.5$\times 10^{4}\mathrm{M_{\odot}}$ \& 4.5$\times 10^{5}\mathrm{M_{\odot}}$, respectively. The Plummer equivalent gravitational softening of 
the collisionless components, i.e., stars, dark matter, is 288 parsecs at z = 0. For gas, the gravitational softening is adaptive, and its minimum is kept at 74 comoving parsecs at all times.

Haloes in TNG simulations are identified using a FoF algorithm, while galaxies (Subhaloes) residing in a halo are identified using the SUBFIND algorithm \citep{Springeletal2001}. The volume and the resolution of TNG50-1 is ideal to detect the extended low surface brightness disks in galaxies out of a statistical sample of massive galaxies.

\section{Method}
\subsection{Sample selection for massive disk galaxies}
We select all the galaxies (subhaloes) with total stellar mass (defined as the sum of the mass of all star particles gravitationally bound to the subhaloes) $>=10^{11}M_{\odot}$ at z=0, i.e., from snapshot 99 of TNG50-1, and thus obtain an initial parent sample of 132 massive galaxies. We note that we select this mass cut-off based on the reasons discussed in Section 1. Thus, all the double-disk galaxies found in this work will lie above this mass cut-off.

We use the synthetic, idealized SDSS (Sloan Digital Sky Survey \citep{Yorketal2000})-g,r band images of galaxies, made using radiative transfer code SKIRT and available in the TNG supplementary data catalog \citep{Gomezetal2019}, for the morphological measurements. We note that these images in the catalog, corresponding to z=0, were created by assuming that the galaxies are actually located at z=0.0485, as would be observed by a local observer. The number of pixels in each image is chosen such that the resulting pixel scale matches that of the observations, i.e., 0.396 arcsec pixel$^{-1}$ at z=0.0485. This corresponds to a linear scale of 0.276 $h^{-1}$ckpc (ckpc refers to comoving kpc) or 0.388 kpc in physical unit, as mentioned in the details of this data catalog. We note that the unit of each image is in electron counts $\mathrm{s^{-1}pixel^{-1}}$. The magnitude photometric zero-points in AB magnitude system are calculated using $2.5\mathrm{log10(fluxmag0)}$, where fluxmag0 for SDSS-g, r bands are $2.4\times 10^{10}$, and $2.38\times 10^{10}$ electron counts $\mathrm{s^{-1}}$, respectively (as given on the TNG website). By construction, each synthetic image contains one galaxy at a random orientation. The field of view is equal to 15 times the 3D stellar half-mass radius of the corresponding galaxy \citep{Gomezetal2019}. We note that these are idealized images as would be hypothetically observed by an instrument with a point-like PSF (i.e., a Dirac delta function), and an infinite signal-to-noise ratio. We do not add any noise or sky background to these images or convolve them with a PSF. We use them in their idealized form only. This makes the images most suitable to detect the extended, low surface brightness outskirt of the galaxies \citep[see][]{Zhuetal2023}. Observational realism has been found not to have a significant effect on the values of the best-fit parameters of a model (see \citet{LaChanceetal2024}).

First, we visually inspect each galaxy image to identify the ones showing the signature of strong tidal disturbances/interactions, jets, etc, in the outskirts.  
For simplicity, we discard these cases (47 in number). Because the above features can make their 2D light modeling and structural decomposition complicated, especially in the galaxy's outskirts. This will make it difficult to determine the presence of an extended faint disk, if present, robustly. It is also hard to model an extended LSB disk, if present, when a galaxy is at a perfect edge-on orientation. Therefore, for simplicity, we discard 13 edge-on cases as well. This gives us a sample of 72 galaxies that contains ellipticals, S0s, and spirals.

Now, in order to determine the disk galaxies from the above sample and discard the ellipticals, we perform 2D GALFIT modeling \citep{Pengetal2002,Pengetal2010} of each galaxy. We fit a single S\'{e}rsic model, using GALFIT, on the synthetic g-band images of the galaxies. The output of the GALFIT modeling consists of the best-fit S\'{e}rsic model parameters (S\'{e}rsic index "n", effective radius $R_{e}$, axis ratio, position angle, and center coordinates). 
We consider the galaxies with S\'{e}rsic index n$<$2.5 to be disk galaxies and the galaxies with n$>$2.5 to be ellipticals \citep{Shen2003, Lange2015, Sachdeva2020}. We also checked that in some cases the single S\'{e}rsic modeling spuriously gives n$>$2.5 for galaxies that contain a prominent spiral structure or ring, leading them to be misclassified as ellipticals in the above process. We show the distribution of the S\'{e}rsic index "n" in Figure \ref{fig:fig1} except one case where the GALFIT result did not converge. It was visually classified as a disk galaxy. Therefore, we finalize the number of disk galaxies to be 57, and elliptical galaxies to be 15, based on GALFIT modeling as well as visual inspection of the galaxy images wherever necessary. We used this sample of 57 disk galaxies to find the ones that have a double-exponential disk structure.

We note that this parent sample of 57 from 132 is a lower limit of disk galaxies as the actual number is higher, as evident from our selection process. The aim of the selection process was to ensure a robust detection of double-exponential disk galaxies subsequently. Thus, the double-disk galaxies detected will be a lower limit w.r.t all the 132 massive galaxies from TNG50.

\begin{figure}
\centering
\includegraphics[width = 0.42\textwidth, height=0.30\textwidth]{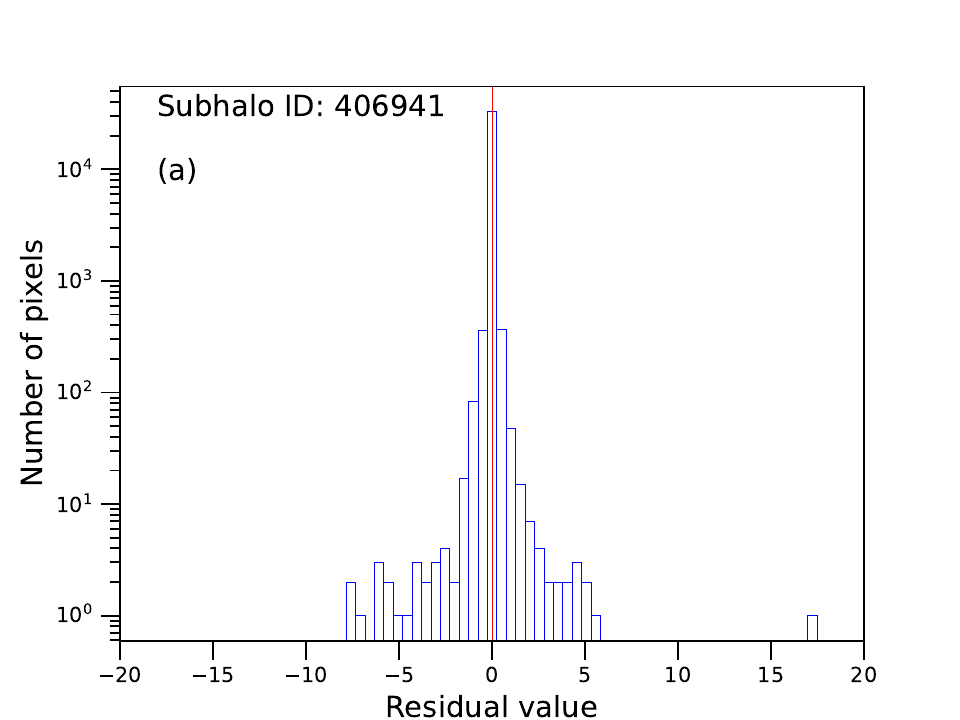}
\includegraphics[width = 0.42\textwidth, height=0.30\textwidth]{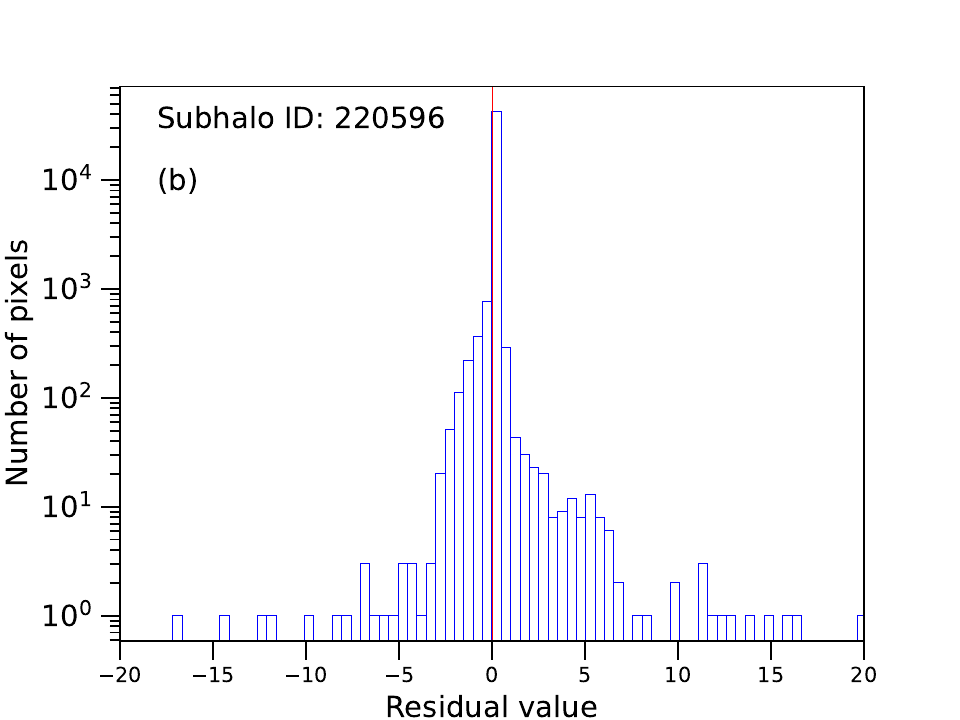}
\includegraphics[width = 0.42\textwidth, height=0.30\textwidth]{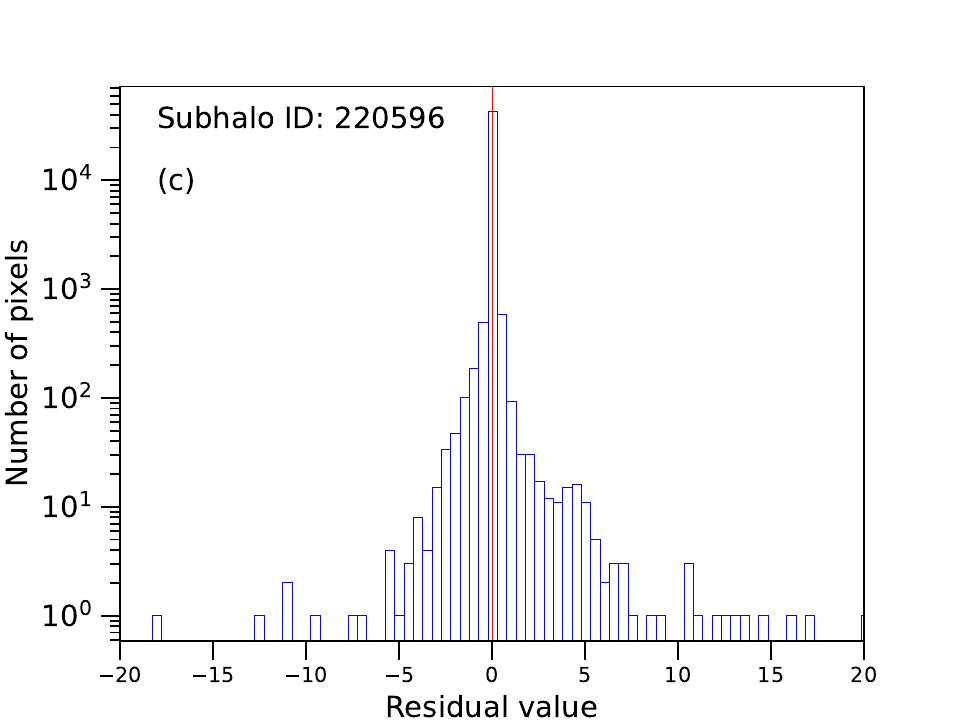}
\caption{Histogram of distribution of GALFIT residual for two galaxies in the synthetic SDSS-g band. For the purpose of comparison, the range of the x-axis is taken to be the same in each case. The red vertical line, passing through the zero value of residual, separates the positive and negative residual. \textbf{Panel (a)}: showing a symmetric histogram for subhalo ID:406941 fitted with S\'{e}rsic plus exponential. \textbf{Panel (b)}: The same for subhalo ID: 220596, which reveals positive skewness - a case for the double disk model. \textbf{Panel (c)}: Same as in panel (b) but with a S\'{e}rsic plus two exponential disks. The histogram here clearly appears to be more symmetric upon the addition of an extended low surface brightness disk.}
\label{fig:fig2}
\end{figure}

\subsection{Finding the double-exponential disk galaxies through GALFIT modeling}

\begin{figure*}
\centering
\includegraphics[width = 0.33\textwidth, height=0.25\textwidth]{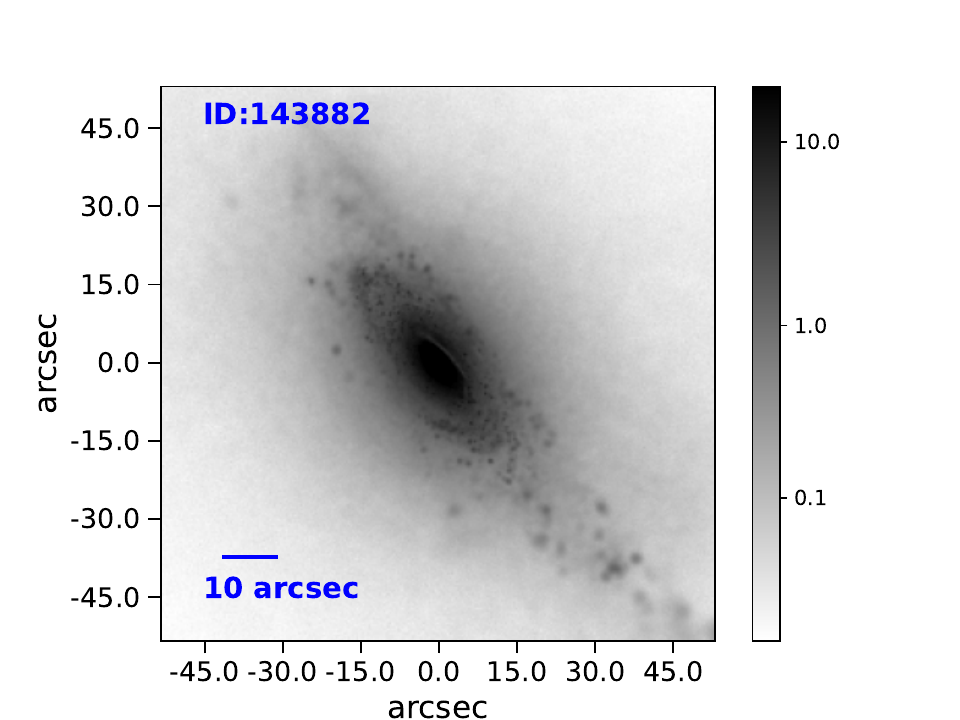}
\includegraphics[width = 0.31\textwidth, height=0.25\textwidth]{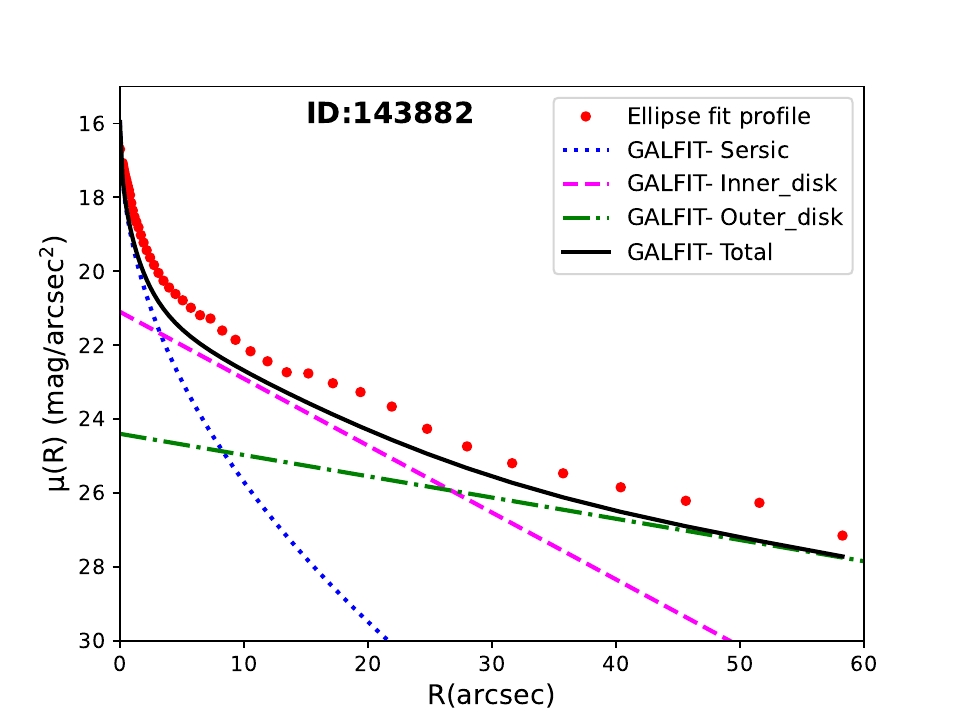}
\includegraphics[width = 0.33\textwidth, height=0.25\textwidth]{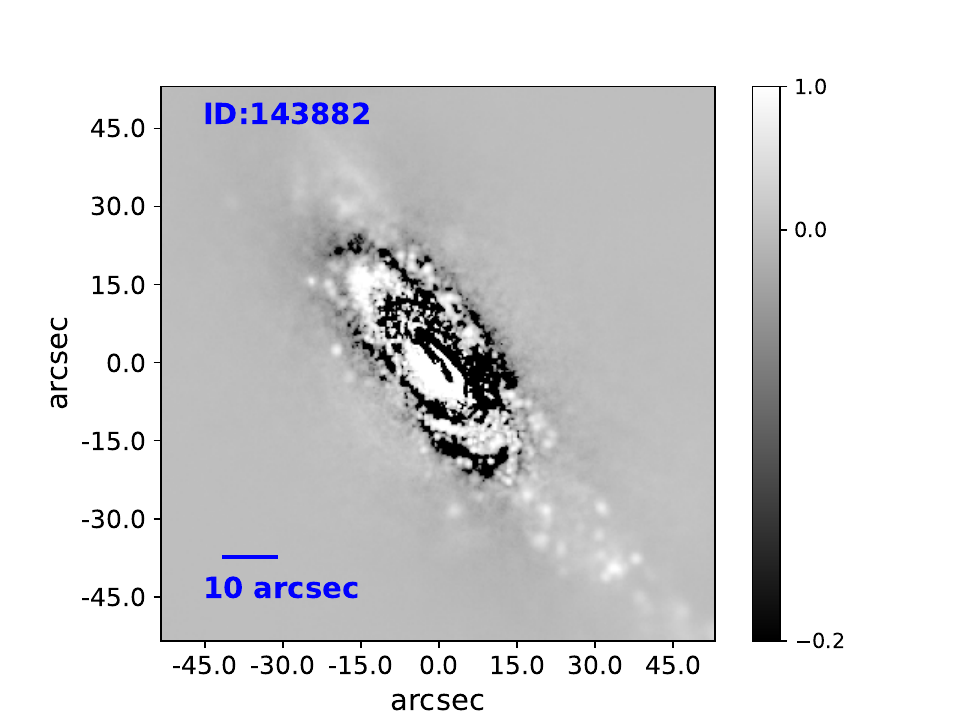}
\medskip
\includegraphics[width = 0.33\textwidth, height=0.25\textwidth]{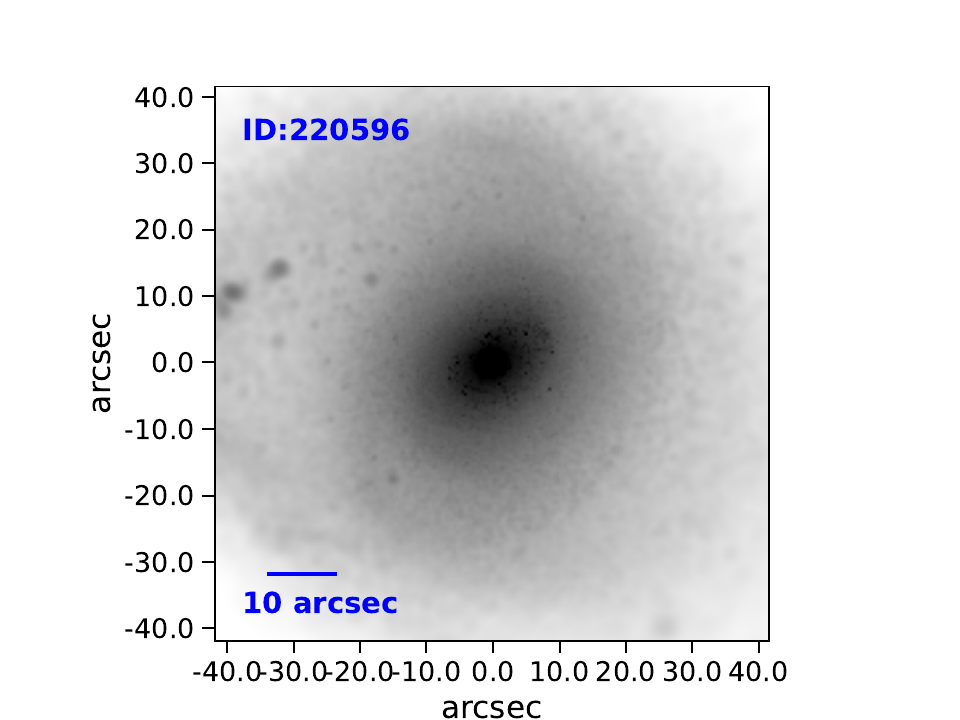}
\includegraphics[width = 0.31\textwidth, height=0.25\textwidth]{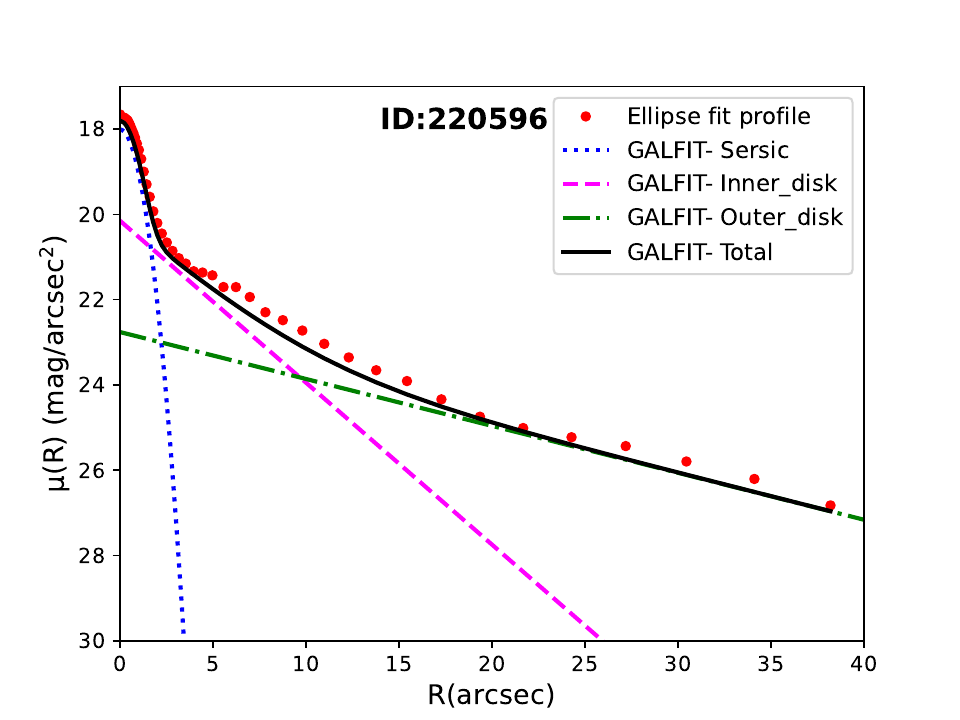}
\includegraphics[width = 0.33\textwidth, height=0.25\textwidth]{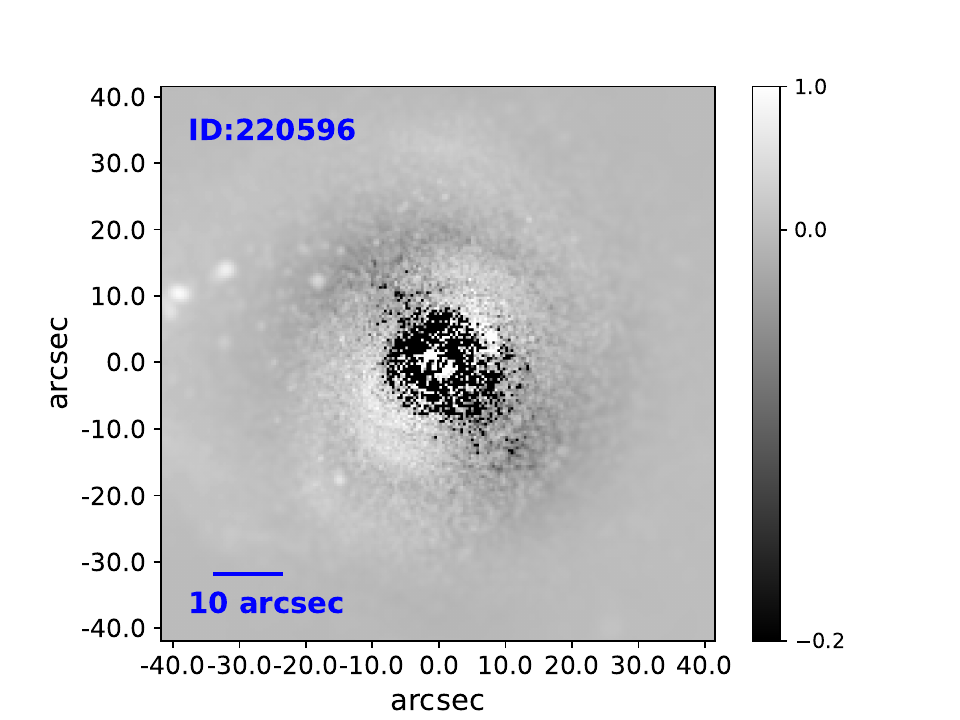}
\medskip
\includegraphics[width = 0.33\textwidth, height=0.25\textwidth]{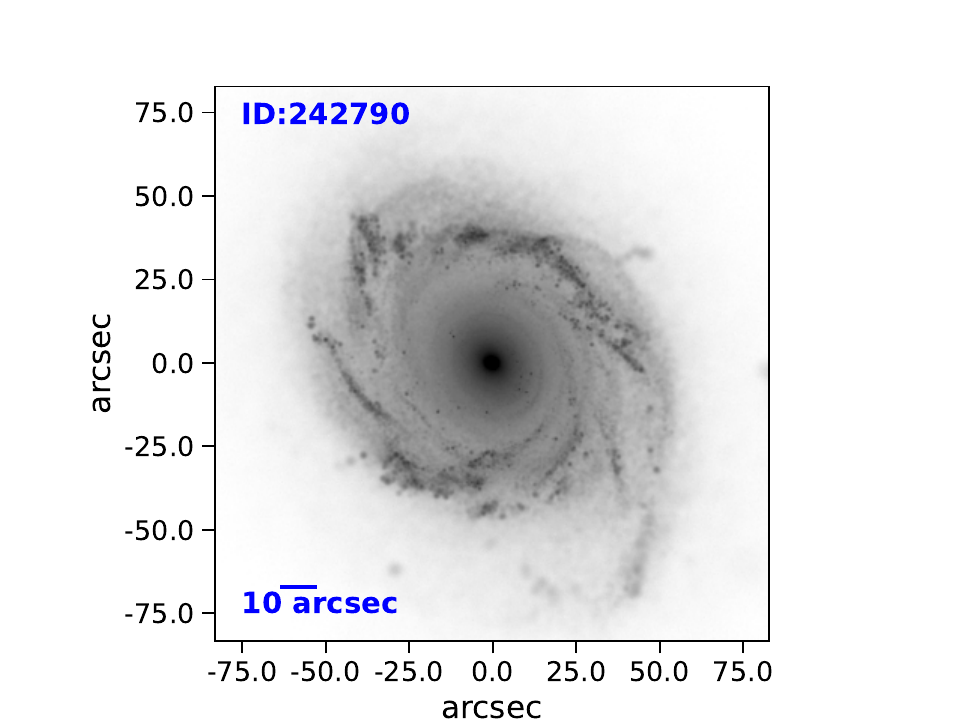}
\includegraphics[width = 0.31\textwidth, height=0.25\textwidth]{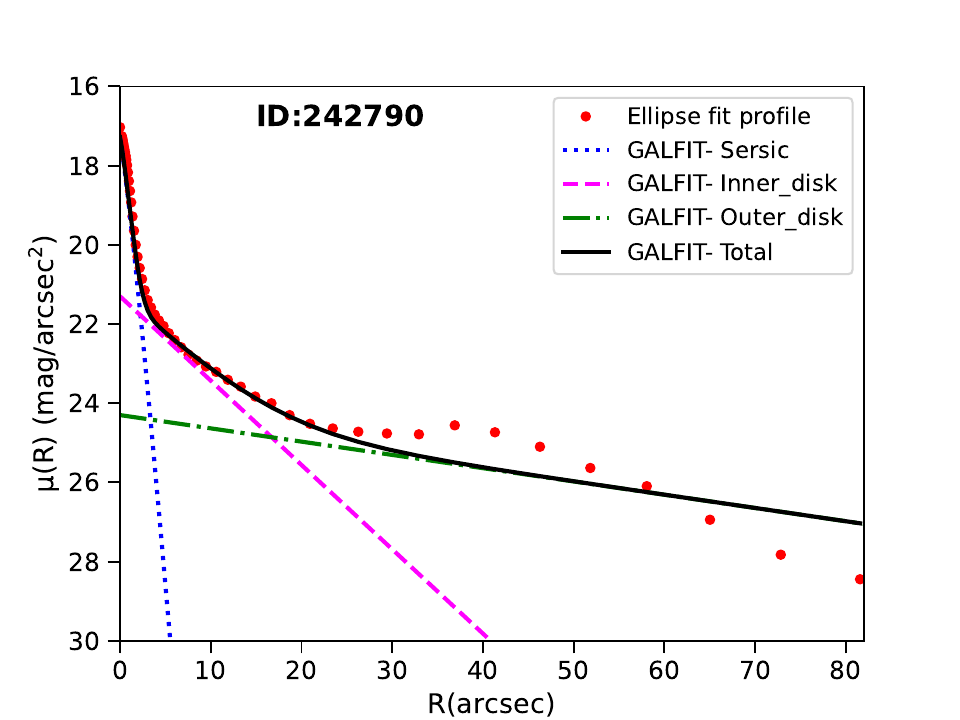}
\includegraphics[width = 0.33\textwidth, height=0.25\textwidth]{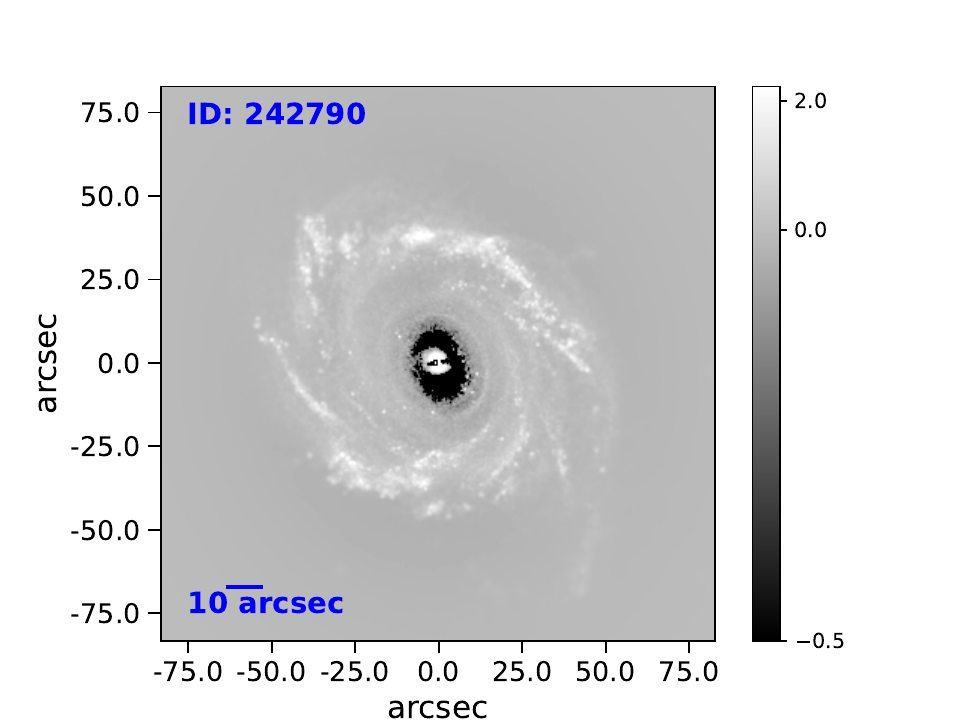}
\medskip
\includegraphics[width = 0.33\textwidth, height=0.25\textwidth]{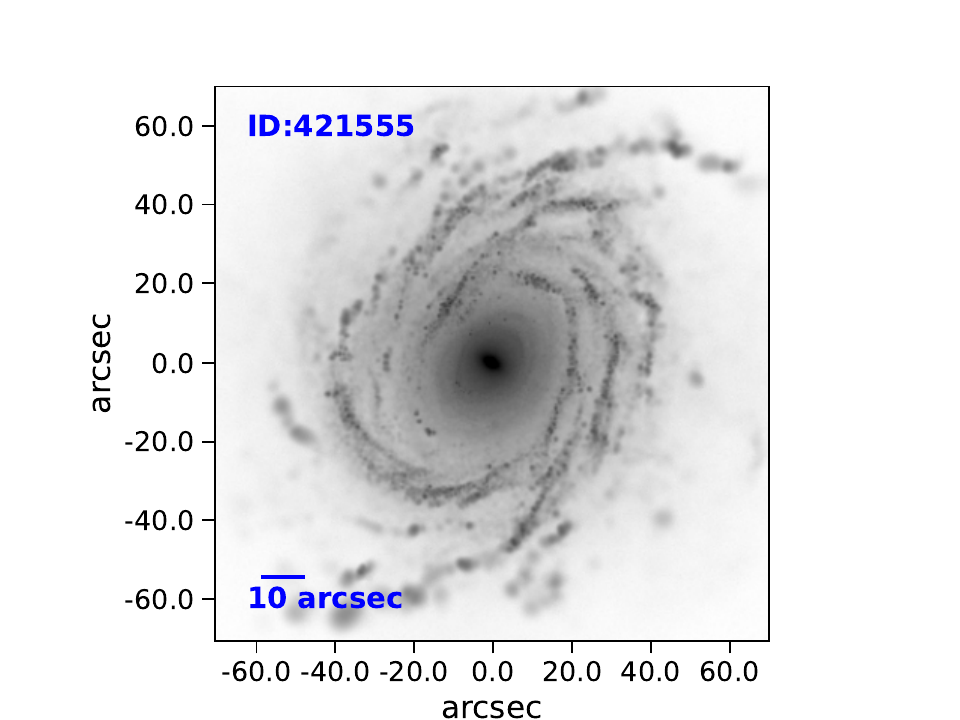}
\includegraphics[width = 0.31\textwidth, height=0.25\textwidth]{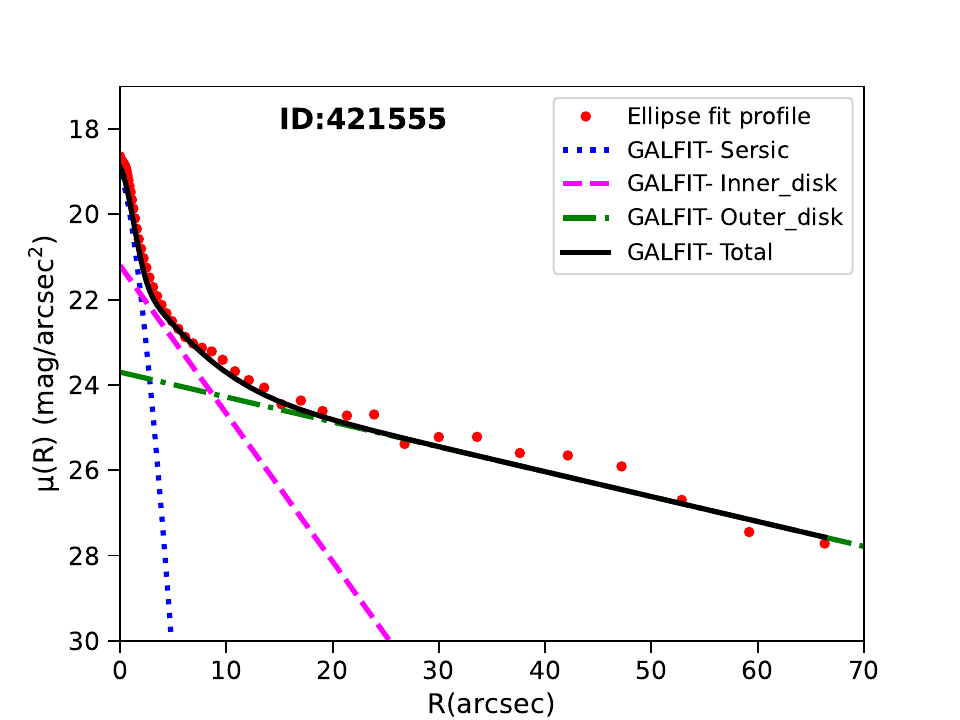}
\includegraphics[width = 0.33\textwidth, height=0.25\textwidth]{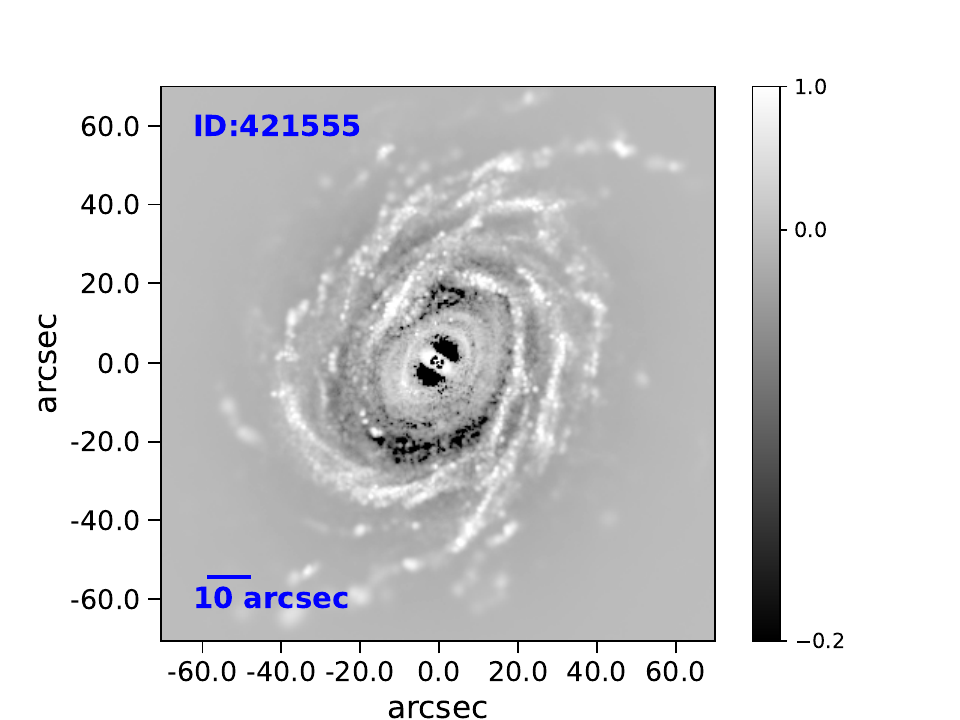}
\medskip
\caption{Simulated galaxies (subhaloes) with double-exponential i.e., HSB plus LSB disk structures. The subhalo IDs are mentioned in each case. All the images are displayed in the image unit. \textbf{Left column}: Synthetic SDSS g-band idealized images, shown using the same color bar (in log stretch). \textbf{Middle column}: Plot of the 1D surface brightness profiles (red data points) obtained from elliptical isophote fitting of the input images. The central S\'{e}rsic (blue dotted curve), inner (magenta dashed curve), and outer exponential (green dash-dotted curve) disk profiles, calculated analytically from the corresponding best-fit 2D GALFIT model of the image, are over-plotted on the above 1-D surface brightness profiles. The total, i.e., the analytical sum of the GALFIT model profiles is denoted by the black solid curve. \textbf{Right column}: Residual images obtained from GALFIT modeling displayed following the color bar (in log stretch) provided beside each.}
\label{fig:fig3}
\end{figure*}

\begin{figure*}
\vspace{-80mm}
\centering
\includegraphics[width = 0.33\textwidth, height=0.25\textwidth]{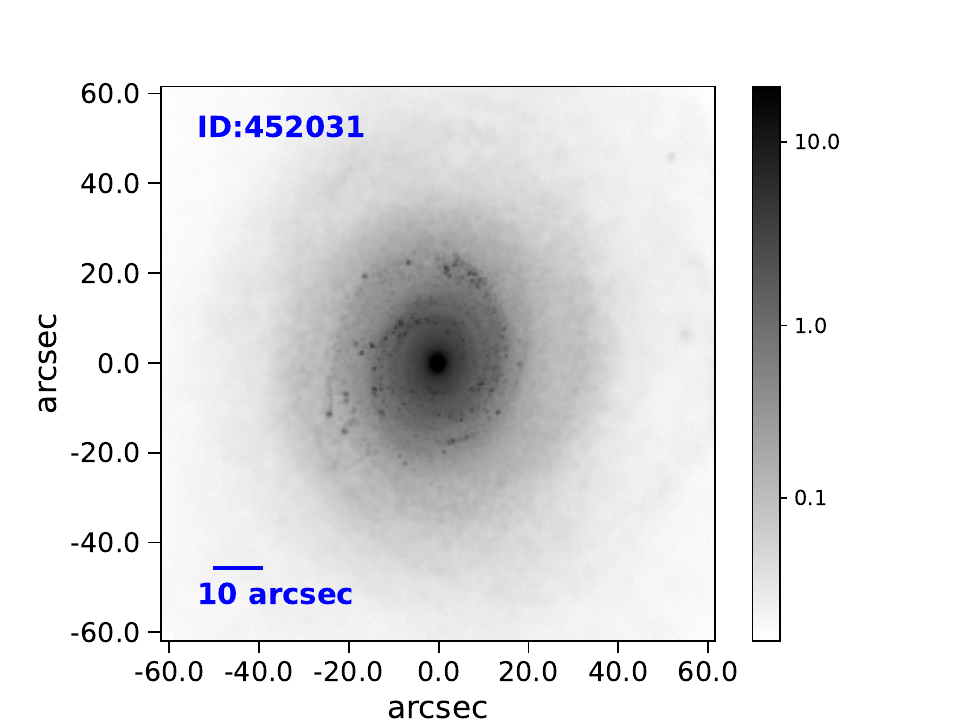}
\includegraphics[width = 0.31\textwidth, height=0.25\textwidth]{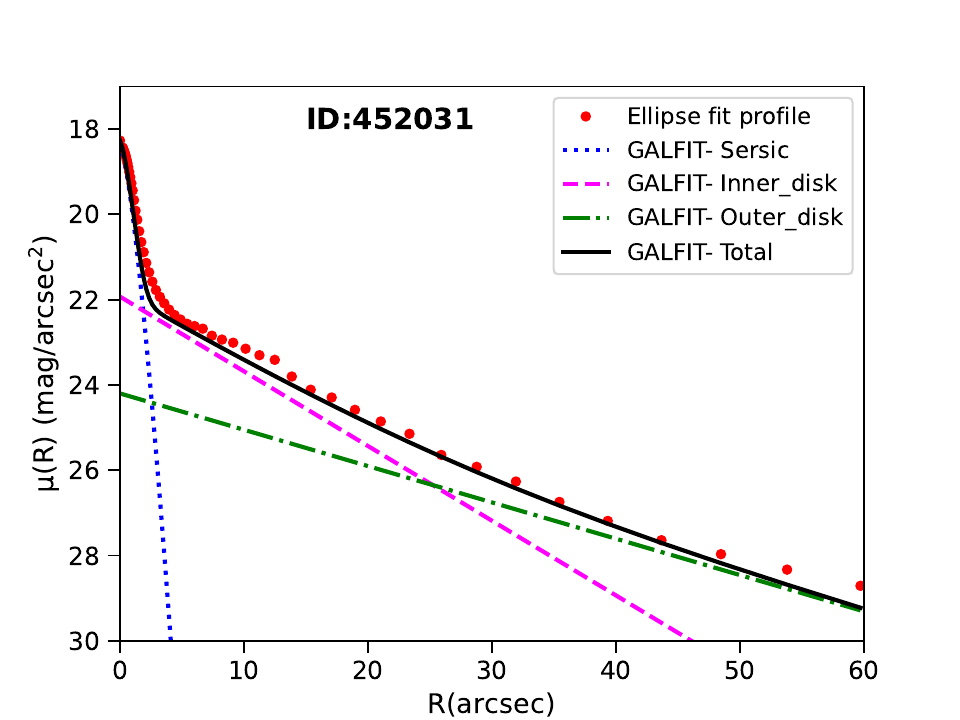}
\includegraphics[width = 0.33\textwidth, height=0.25\textwidth]{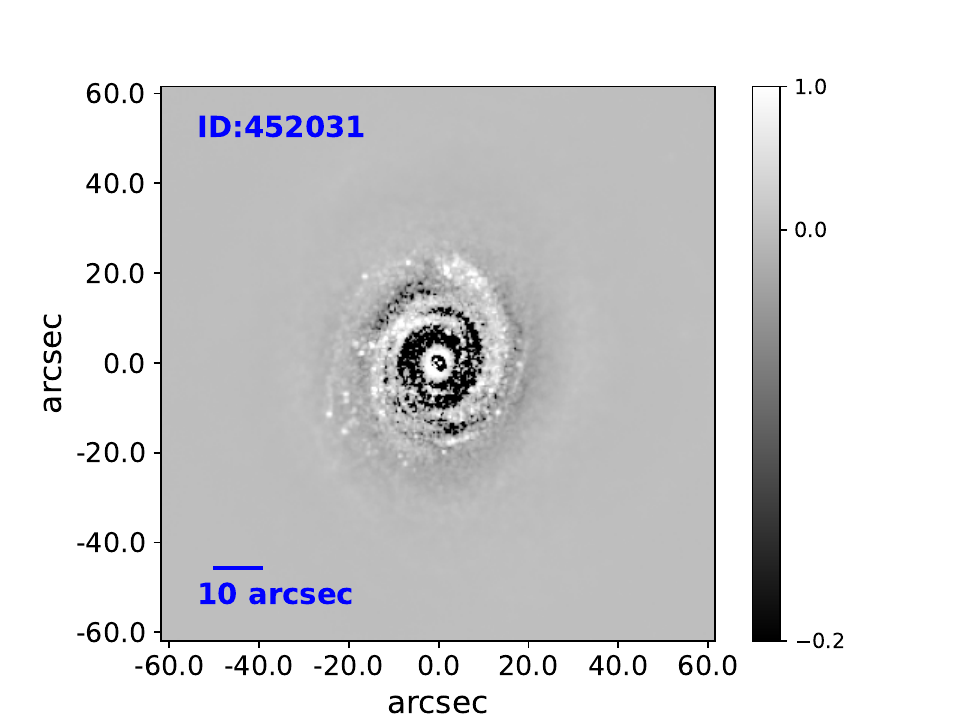}
\medskip
\includegraphics[width = 0.33\textwidth, height=0.25\textwidth]{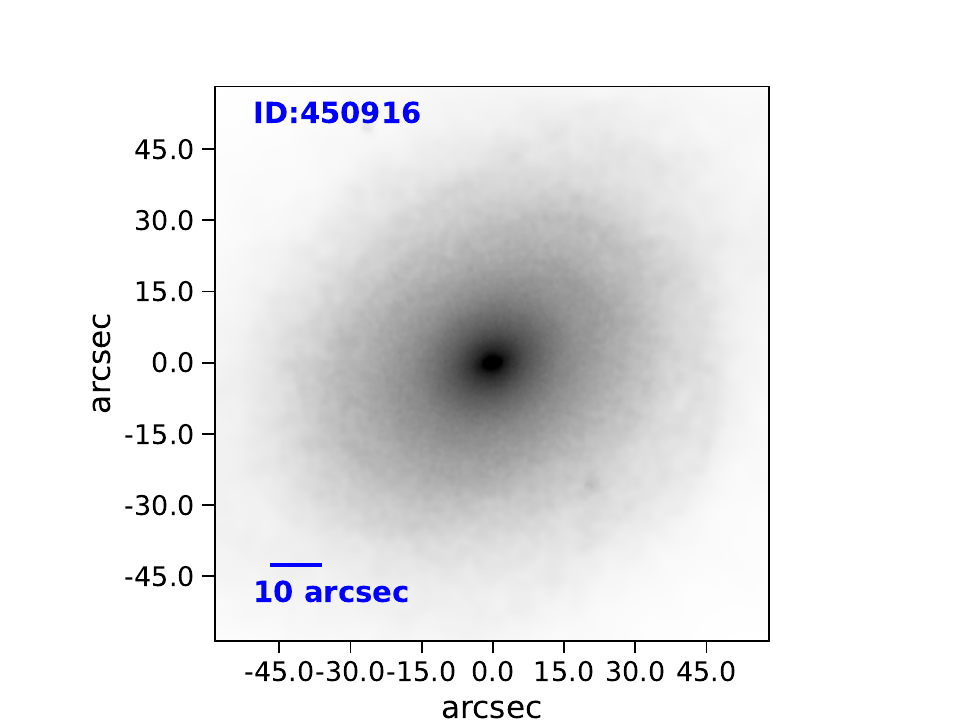}
\includegraphics[width = 0.31\textwidth, height=0.25\textwidth]{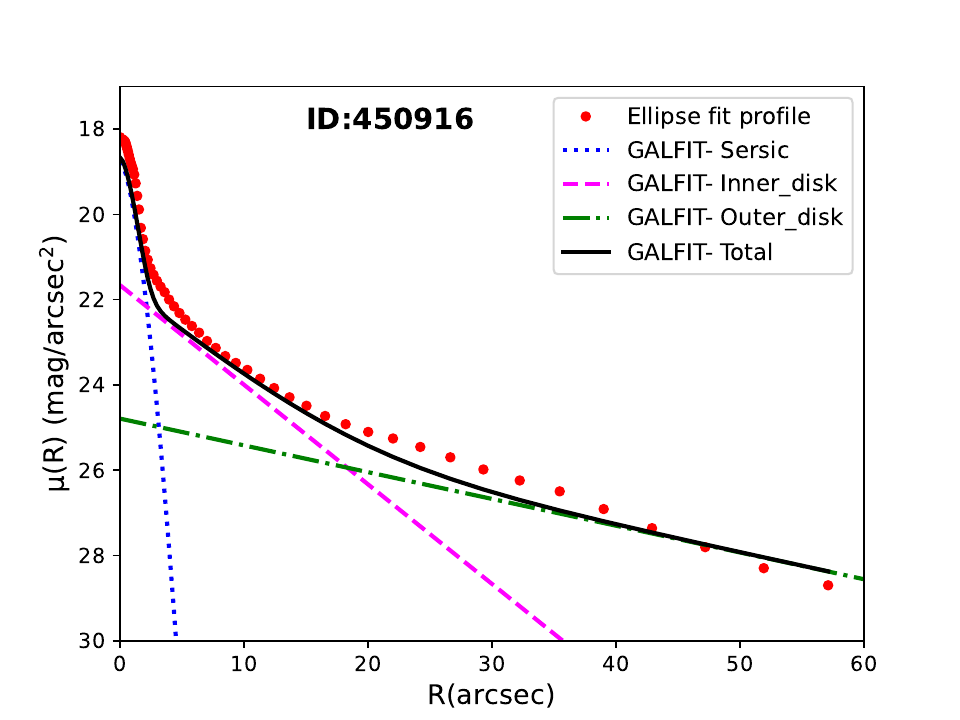}
\includegraphics[width = 0.33\textwidth, height=0.25\textwidth]{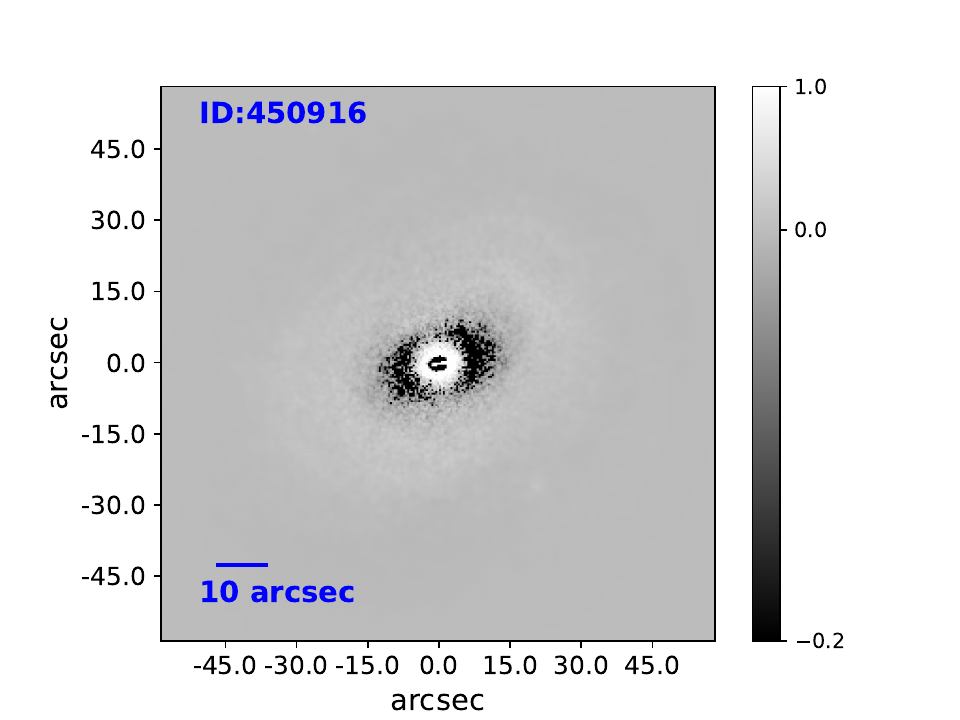}
\medskip
\includegraphics[width = 0.33\textwidth, height=0.25\textwidth]{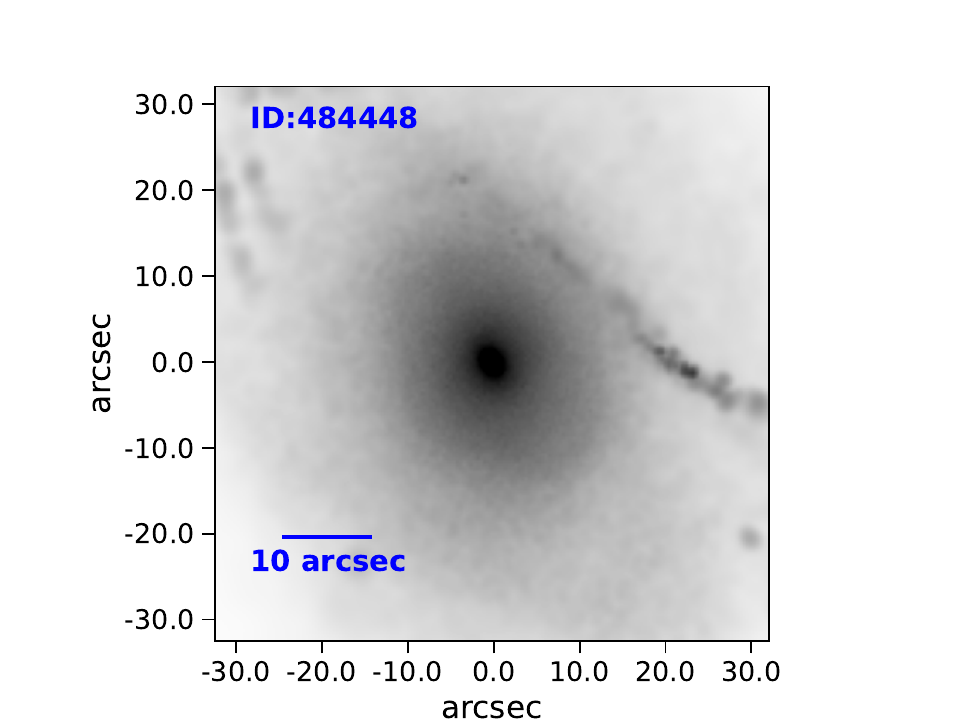}
\includegraphics[width = 0.31\textwidth, height=0.25\textwidth]{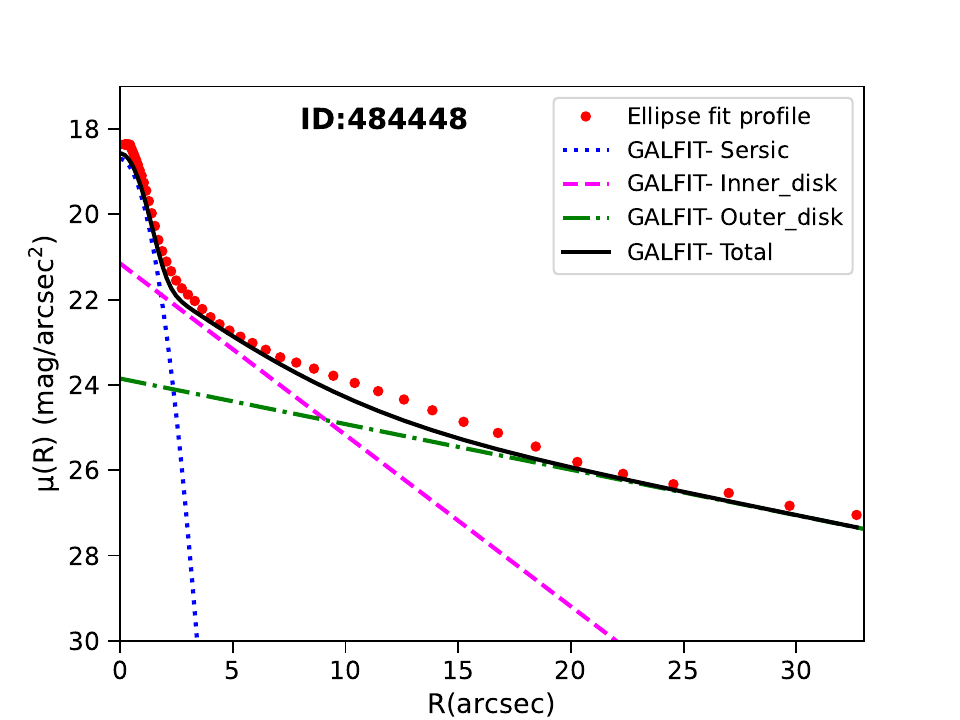}
\includegraphics[width = 0.33\textwidth, height=0.25\textwidth]{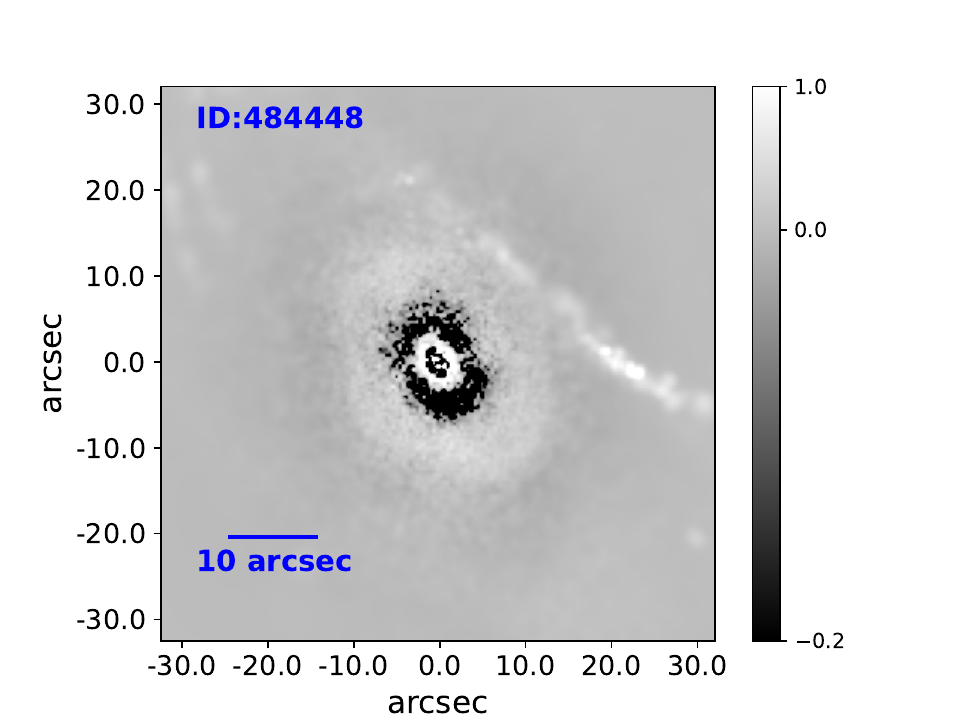}
\medskip
{\textbf{Figure 3.} Continued from the above figure}
\end{figure*}

\noindent
To find out the galaxies that can be best described by the combination of two exponential disks among the sample of 57, first, we perform a S\'{e}rsic plus single exponential disk modeling using GALFIT. The S\'{e}rsic function is used to model the central bulge/bar component, and the exponential function models the disk component. For simplicity, we do not model the spiral structure or star-forming rings. We use the best-fit single S\'{e}rsic model parameters, obtained earlier, as the initial fit parameters for the S\'{e}rsic component. The center coordinates of the S\'{e}rsic and exponentials are kept at the centers obtained from single S\'{e}rsic modeling. The rest of the initial parameters for both components are kept free. After running GALFIT, we obtain the best-fit model image and a residual image, which is basically the model image subtracted from the input galaxy image. 

Now, the galaxies that are well modeled with a S\'{e}rsic plus single exponential disk would produce a residual image with both positive and negative values. But if an outer exponential disk is present in the galaxy, the above model would fail to model the flux in the outskirts of the galaxy. As a result, the residual image will show mostly/systematically positive values in the outskirt of the image. One can visually analyze the residual image of each galaxy to detect such cases. Instead of following this tedious procedure, we make a histogram of the distribution of the residual values. We found that 46 cases show a positively biased histogram with an extended distribution of residual. Therefore, an extra component w.r.t S\'{e}rsic and a single exponential disk is required in this subsample. The rest of the sample of galaxies show almost symmetric histogram and hence we consider a S\'{e}rsic plus single disk model to be a good description of them. We note that this could be a useful, fast technique for exploring a large sample.

In the second step, we fit those 46 galaxies with a second exponential disk. We applied the fitting conditions, i.e., techniques as followed in the previous step. Now we visually inspect all the residual images and compare those with their corresponding single disk residual images. We found seven galaxies out of 46 cases where the residual images corresponding to the single disk model have a systematic positive bias in the outskirt which get removed and the residual images becomes more symmetrized around zero values after the addition of the second disk component. We do not expect any significant change in the central part of the residual image as the second disk component present in the galaxy would make very small change in the central regime due to its faint surface brightness. We also verified from the input image of each of those galaxies that there is indeed a substantial disk-like light distribution in the outskirt. We present the GALFIT modeling of these seven galaxies, a detailed comparison between their single and double-exponential models, goodness or accuracy of the fit, and bring out the physically motivated requirement for the second disk in them in the following Section. We also show the histograms corresponding to a single-disk galaxy and a double-disk galaxy in Figure \ref{fig:fig2} in this Section.

We had found such positive residue in the outskirts of some of the other galaxies out of 46 as well. However, the second disk could not be fitted in them with physically meaningful parameters. It could be fitted only at the expense of shrinking the radial scale length of the inner disk extent (within $\sim$ 1 - 1.5 kpc), much smaller than that corresponding to their single disk model. Therefore, we do not consider these cases to genuinely contain a second outer disk. For all the other galaxies in the sample of 46, the positive residual visible in the histogram was found to be solely arising from a central bar, rings, strong spiral arms, stellar halo etc, and not from an extended disk region. Based on the above analysis, we conclude those seven galaxies to be the double-exponential disk galaxies from our sample of 57. We note that this selection process helped us to get a clear, robust sample of such systems. 

We note that these 7 galaxies consist of 12\% of the parent disk galaxy sample (57) used from a simulation volume of $\sim$ 50$^{3}$ Mpc$^{3}$.

\section{Results}
\subsection{The double-exponential disk galaxies}
We present the GALFIT modeling results obtained for the double-disk galaxies in Figure \ref{fig:fig3}. The input images (corresponding to synthetic SDSS g-band) are shown in the leftmost column, and the residual images are shown in the rightmost column. We have not modeled the spiral structure. Therefore, the spiral structures are clearly present in the residual images. 

In the middle column, we show the 1-D surface brightness profiles (red data points in Fig.3) obtained by fitting elliptical isophotes on the images by running IRAF ELLIPSE task \citep{Jedrzejewski1987} as routinely done in the literature. This basically represents the radial profile of the azimuthal average of the flux of the galaxy. The best-fit central S\'{e}rsic, inner \& outer exponential disk profiles as obtained from GALFIT (the corresponding analytical profile calculations are described in the following subsection) are over-plotted on those 1-D surface brightness profiles. We note that the total, i.e., the analytical sum (black curve) of those component profiles overall reproduces the 1-D surface brightness profile well except in IDs: 143882, 242790. We note that ID:143882 is an inclined disk galaxy. However, the analytical radial surface brightness profile expressions of the over-plotted S\'{e}rsic, exponential components do not take into account the effect of a projected light distribution. Therefore, we find a consistent difference between the original 1-D profile and the total modeled profile in this case. In other words, the above difference does not arise due to an inaccurate GALFIT modeling. We checked that this does not affect the analysis of the necessity of the second disk here. The rest of the sample of galaxies are at a very low inclination and, therefore, are not affected. We note, that there are very strong spiral arms in the LSB outskirt of ID:242790, which affects the fitting of the LSB disk. However, we have checked that this does not affect the conclusion on the presence of the outer LSB disk there. 

\begin{figure*}
\centering
\includegraphics[width = 0.37\textwidth, height=0.27\textwidth]{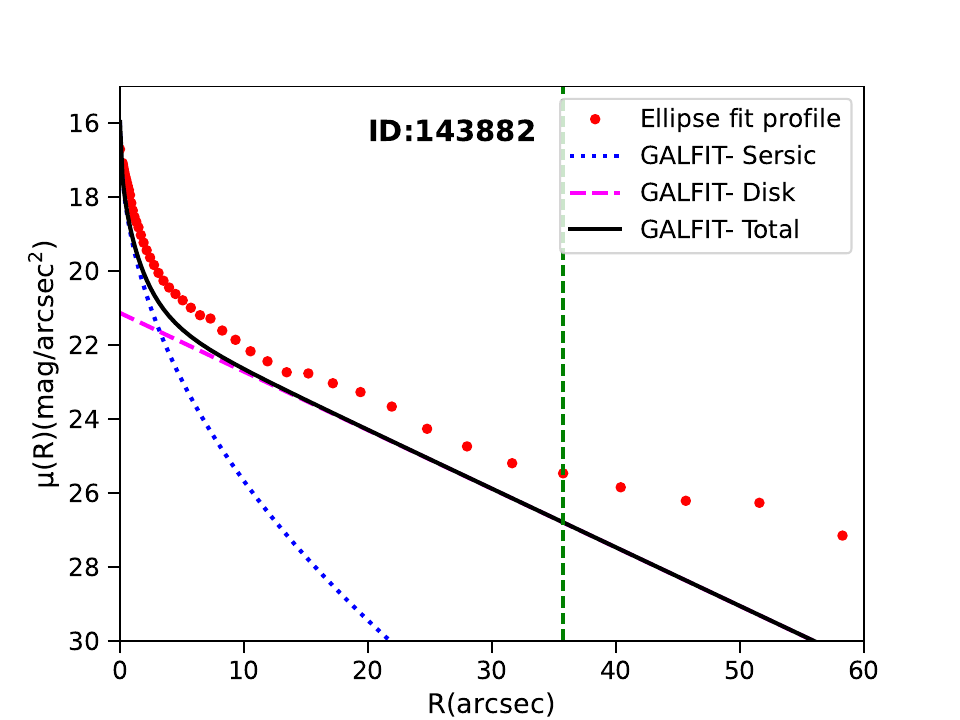}
\includegraphics[width = 0.37\textwidth, height=0.27\textwidth]{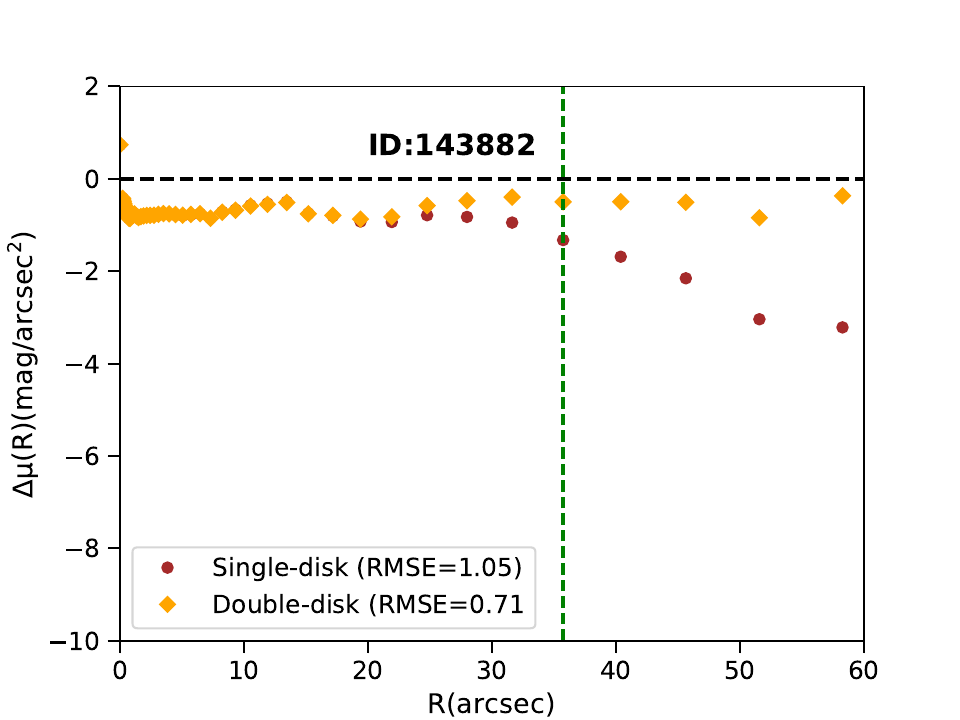}
\medskip
\includegraphics[width = 0.37\textwidth, height=0.27\textwidth]{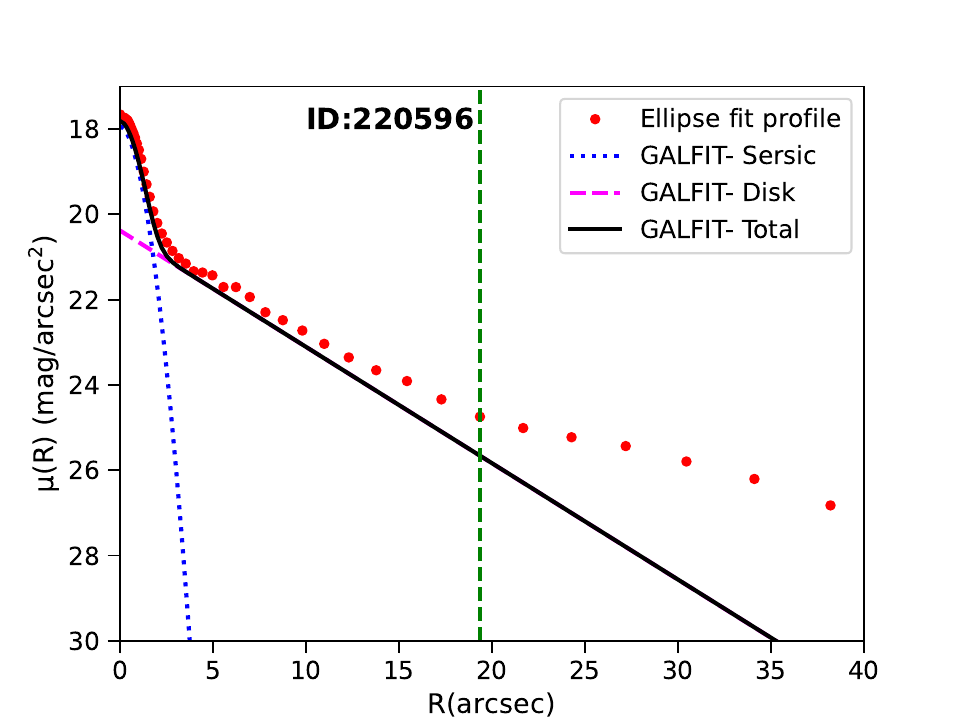}
\includegraphics[width = 0.37\textwidth, height=0.27\textwidth]{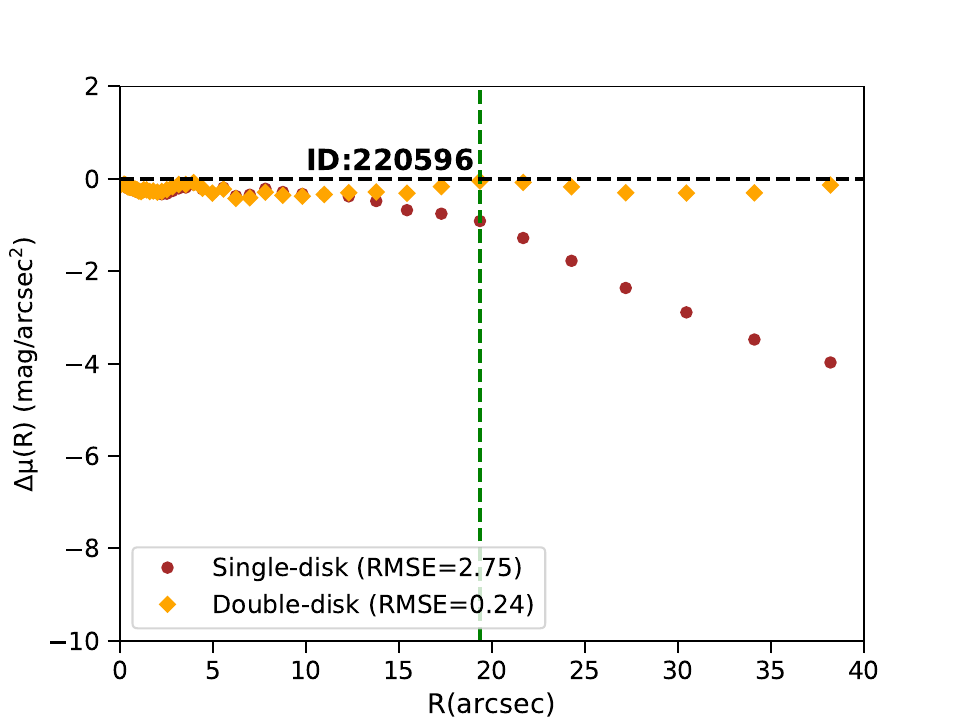}
\medskip
\includegraphics[width = 0.37\textwidth, height=0.27\textwidth]{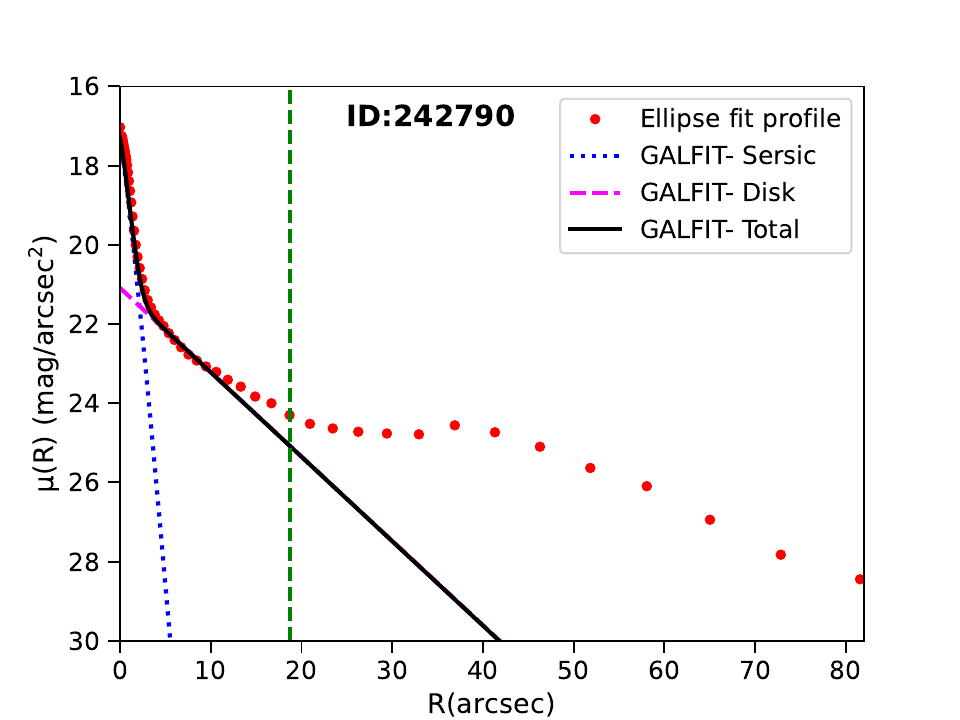}
\includegraphics[width = 0.37\textwidth, height=0.27\textwidth]{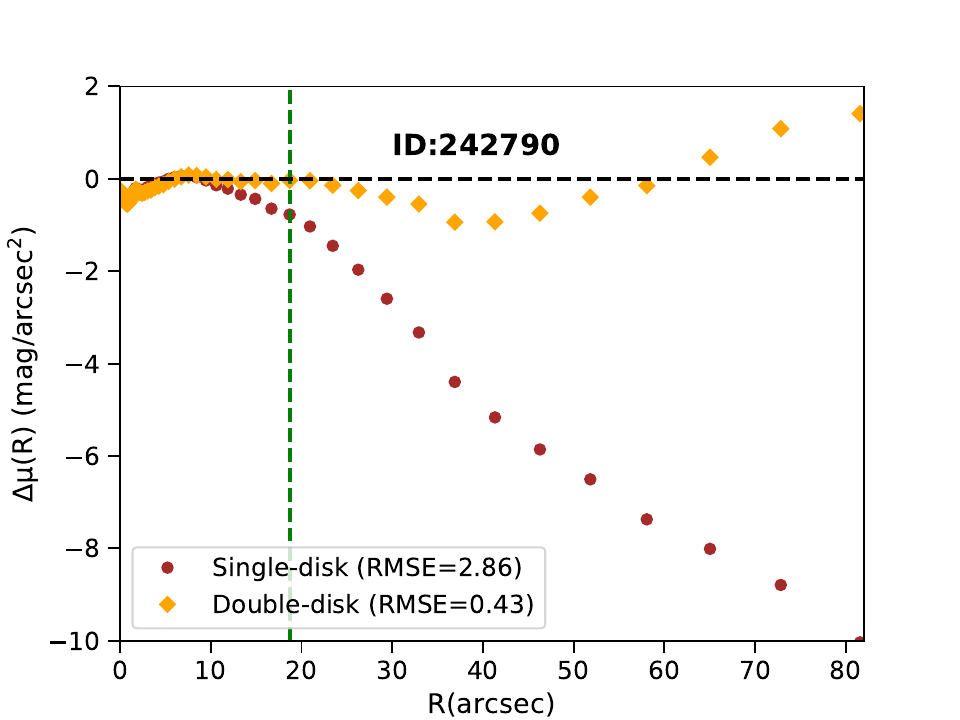}
\medskip
\includegraphics[width = 0.37\textwidth, height=0.27\textwidth]{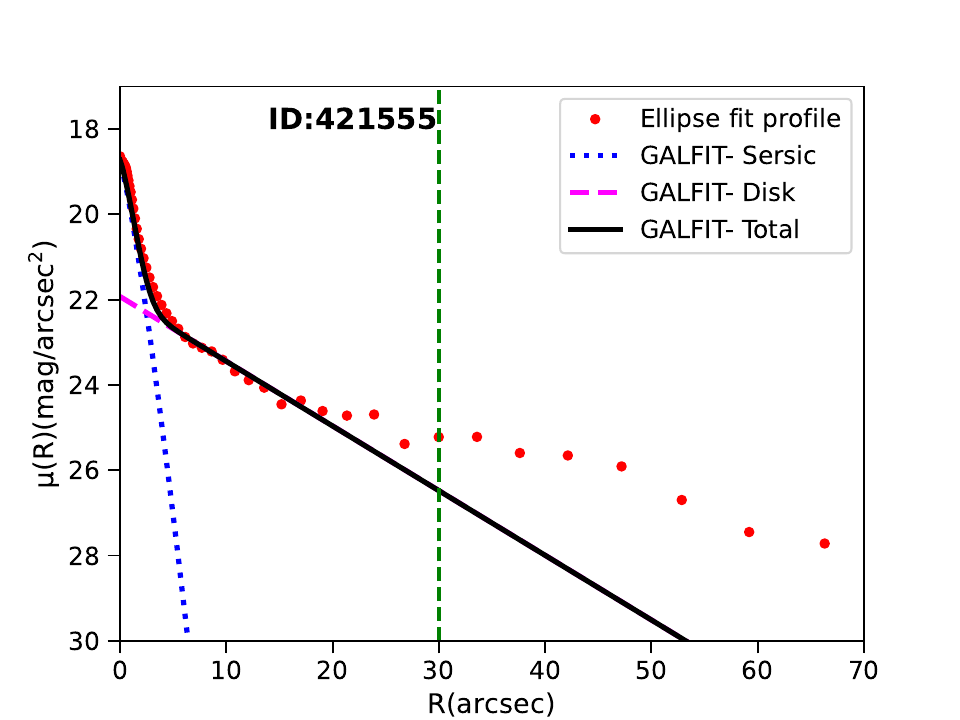}
\includegraphics[width = 0.37\textwidth, height=0.27\textwidth]{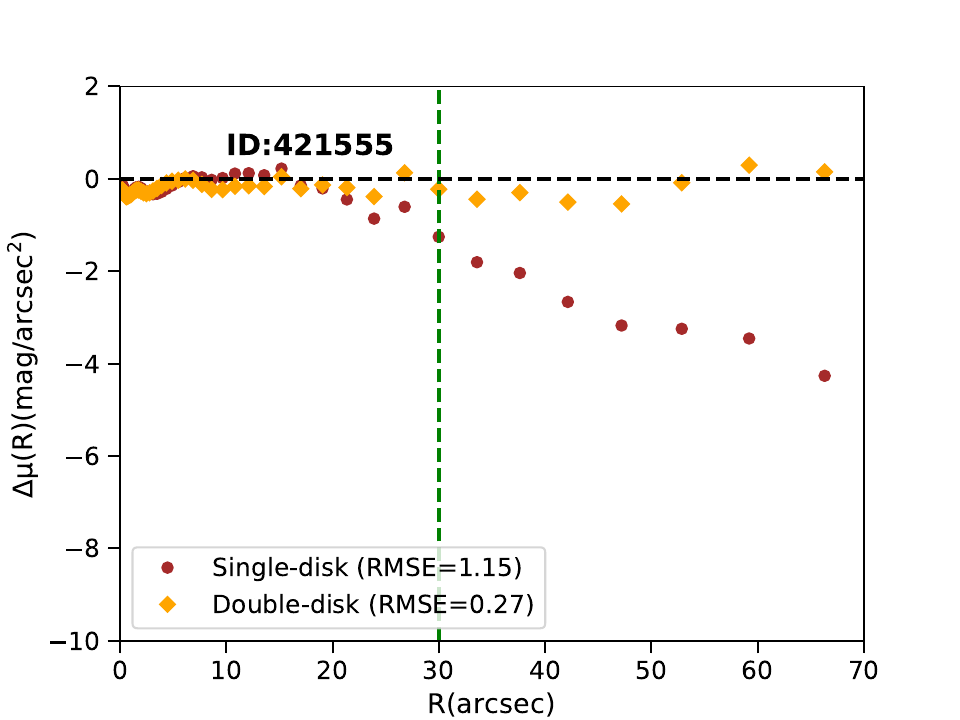}
\medskip
\caption{\textbf{Left column}: The S\'{e}rsic plus single exponential disk modeling of each of the double-disk galaxies. Plot of the 1-D surface brightness profiles (red data points) obtained from elliptical isophote fitting of the input images. The central S\'{e}rsic (blue dotted curve) \& single exponential profile (magenta dashed curve), calculated analytically from the corresponding S\'{e}rsic plus single disk  GALFIT model of the image, are over-plotted on the 1-D surface brightness profile. The black curve is the total, i.e., the analytical sum of the S\'{e}rsic and exponential profiles. The green dashed curve denotes the radial distance from where the slope of the 1-D surface brightness profile changes w.r.t the modeled profile. \textbf{Right column}: The radial variation of the residue that represents the difference between the total galaxy surface brightness profile (red data points) and the modeled surface brightness profile (black curve). The brown circles represent the residue corresponding to the single-disk model, and the orange diamonds represent the residue corresponding to the double-disk model. The root mean square error (RMSE) corresponding to each case is also mentioned.}
\label{fig:fig4}
\end{figure*}
\noindent 

\begin{figure*}
\vspace{-60mm}
\centering
\includegraphics[width = 0.37\textwidth, height=0.27\textwidth]{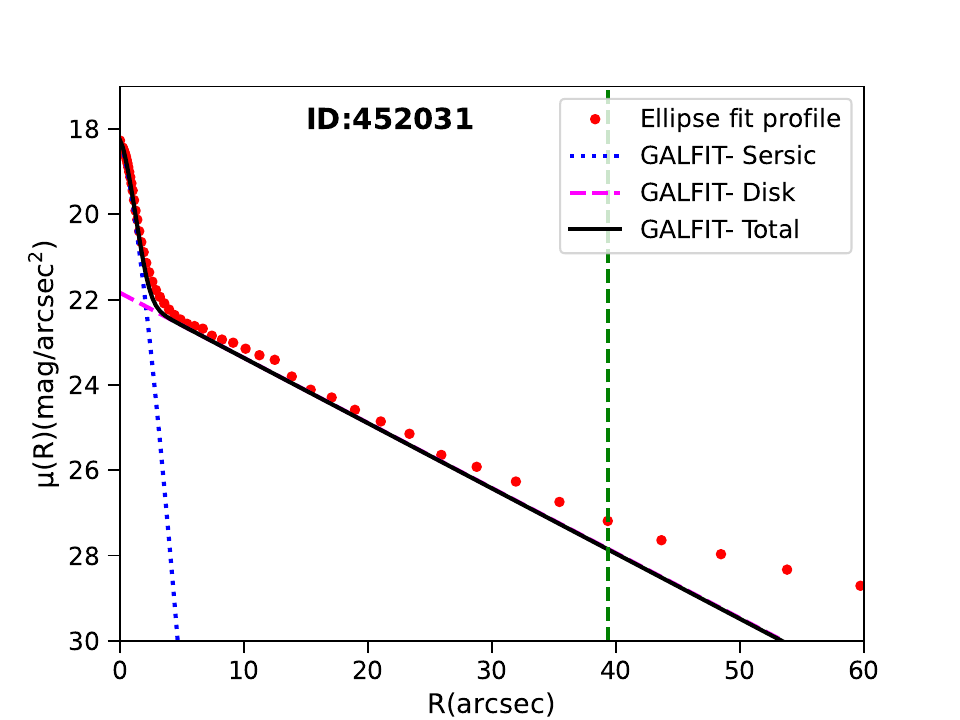}
\includegraphics[width = 0.37\textwidth, height=0.27\textwidth]{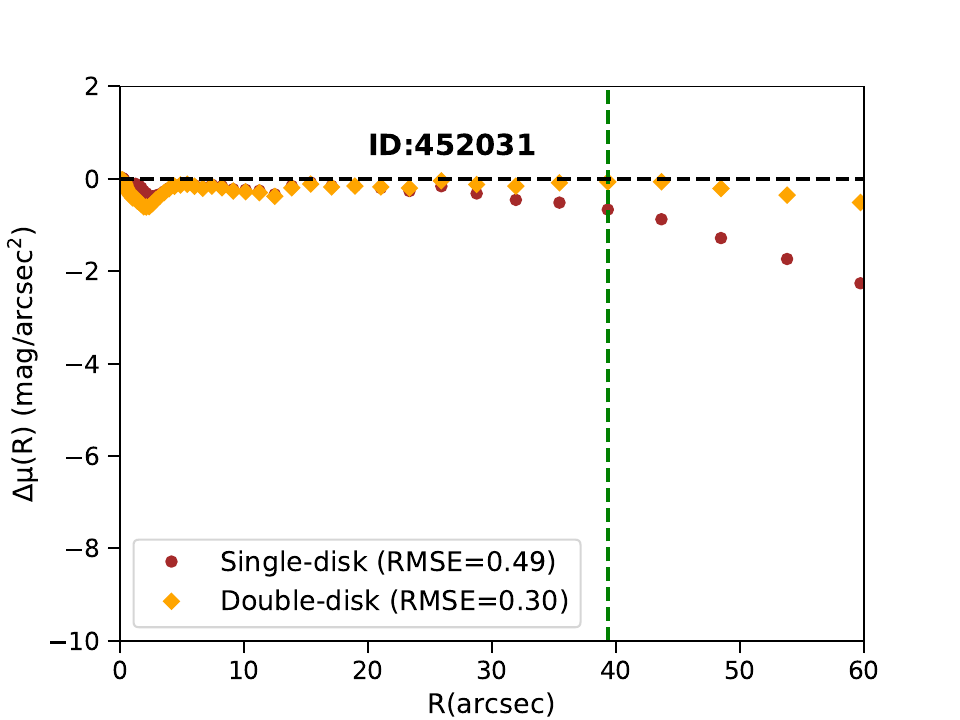}
\medskip
\includegraphics[width = 0.37\textwidth, height=0.27\textwidth]{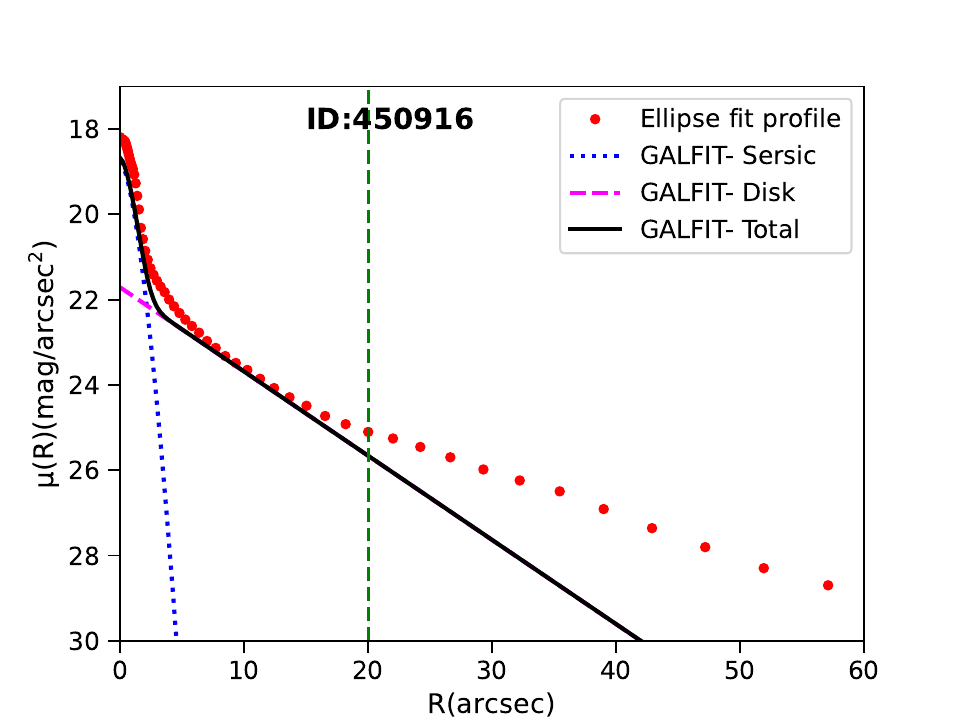}
\includegraphics[width = 0.37\textwidth, height=0.27\textwidth]{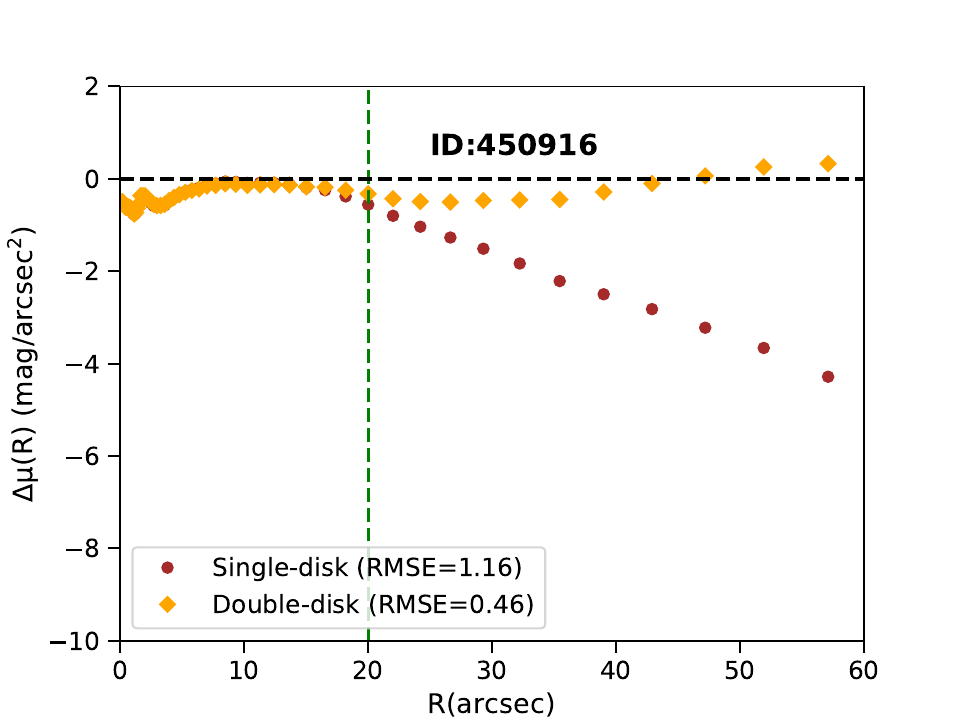}
\medskip
\includegraphics[width = 0.37\textwidth, height=0.27\textwidth]
{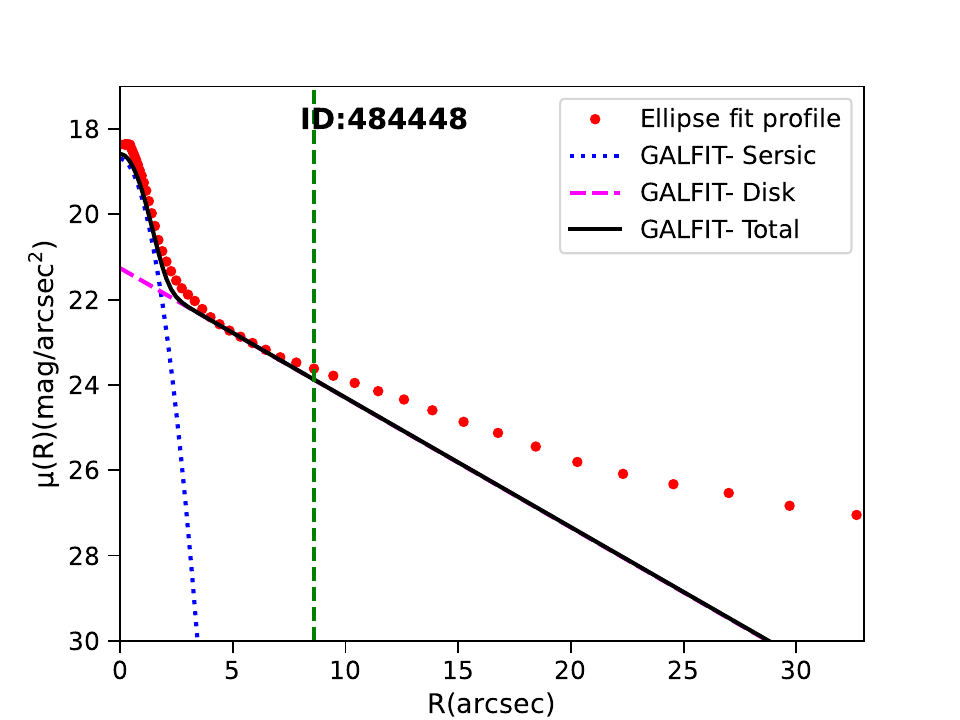}
\includegraphics[width = 0.37\textwidth, height=0.27\textwidth]{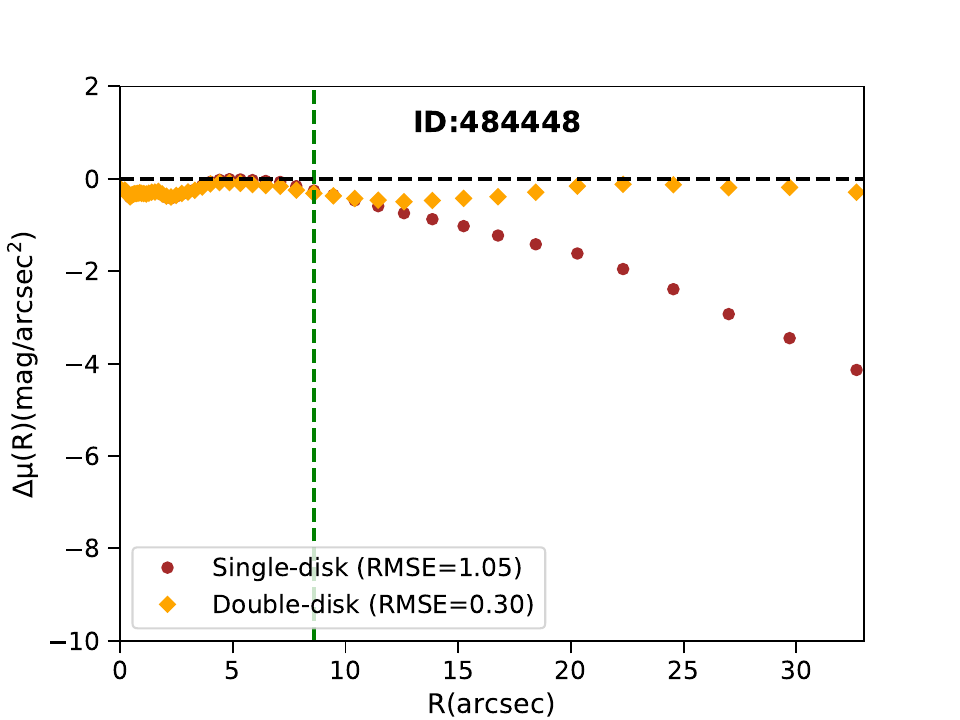}
\end{figure*}

Figure \ref{fig:fig4} shows the validity of the presence of the outer exponential disk for each galaxy. The left column presents the S\'{e}rsic plus single disk model. This clearly shows that a significant flux is left unmodeled in the outskirt of each of them. We denote the radial distance (by a green line in each plot) beyond which the 1-D surface brightness profile (red data points) deviates clearly from the total modeled profile (black curve). In the right column, we plot the difference between the 1-D surface brightness profile and the total modeled profile, corresponding to both single-exponential and double-exponential models. This basically represents the radial profile of residue and, therefore, a measure of the goodness of the fit. Using the above values and the number of data points we also calculated root mean square errors which denote a global measure of the goodness of the fit. For each galaxy, the radial profile of residue improves, i.e., it tends toward zero as well as becomes overall more randomized after the addition of the outer exponential. This, along with the reduced RMSE values, prove that a double-exponential disk model is indeed the physically motivated best description of these seven galaxies. We note that the improvement in the profile is less accurate for IDs: 143882, and 242790 due to the reasons discussed earlier in this section. Nevertheless, Figure \ref{fig:fig4} helps us to conclude the presence of a double-exponential disk structure in these seven galaxies in a robust way.

\subsubsection{Calculation of the structural parameters of the double-disk galaxies}

GALFIT output directly gives us the radial scale length and integrated apparent magnitude of the S\'{e}rsic and the exponential components. 
We use the integrated apparent magnitude \& effective radius of the S\'{e}rsic component ($\mathrm{m_{sersic}}$, $R_{e}$) to further calculate its effective surface brightness ($\mu_{e}$) and central surface brightness ($\mu_{0,sersic}$) values, using the following two equations, respectively. 

\noindent
\begin{equation}
\mu_{e}=m_{sersic}+5\mathrm{log_{10}}R_{e}+2.5\mathrm{log_{10}}\bigg(2\pi n \frac{e^{b_{n}}}{(b_{n})^{2n}}\Gamma(2n)\bigg)
\end{equation}

\noindent

\begin{equation}
\mu_{0,sersic} = \mu_{e}-\frac{2.5\mathrm{b_{n}}}{\mathrm{ln(10)}}
\end{equation}

\noindent where $\mathrm{b_{n}}=1.9992n- 0.3271$ for 0.5$<$n$<$10 \citep{Capaccioli1989, GrahamDriver2005}.

Similarly, we use integrated apparent magnitudes of the disk components ($\mathrm{m_{disk}}$), and their corresponding radial scale lengths ($\mathrm{R_{D}}$) to calculate their central surface brightness values $\mathrm{\mu_{0,disk}}$ as follows.

\begin{equation}
\mu_{0,disk} = m_{\mathrm{disk}} + 2.5\mathrm{log}_{10}(2 \pi \mathrm{R_{D}^{2}})
\end{equation}

Finally, we use the following analytical expressions to calculate the radial surface brightness profiles corresponding to the S\'{e}rsic ($\mu_{sersic}(R)$) and inner \& outer exponential disks ($\mu_{disk}(R)$), respectively, \citep{GrahamDriver2005, GrahamWorley2008}.

\begin{equation}
\mu_{sersic}(R) = \mu_{e}+\frac{2.5\mathrm{b_{n}}}{\mathrm{ln(10)}}\big[(R/R_{e})^{1/n}-1\big]
\end{equation}

\noindent
\begin{equation}
\mu_{disk}(R)=\mu_{0,disk}+1.086\bigg(\frac{R}{R_{D}}\bigg).
\end{equation}

We use the above profiles to over-plot them on the 1-D surface brightness profile for each galaxy in Figure \ref{fig:fig3} as well as in Figure \ref{fig:fig4}. 

\noindent 

We correct $\mu_{0,disk}$ values for inclination using the axis ratio (b/a) of the disk, which is obtained as one of the best-fit parameters from GALFIT output, as well as for cosmological surface brightness dimming effect (since by construction the images include this effect \citep{Gomezetal2019}), using the following equation \citep{Zhongetal2008,PahwaSaha2018}

\begin{equation}
\mu_{0,c} = \mu_{0,disk}+2.5\mathrm{log_{10}(b/a)}-10\mathrm{log_{10}(1+z)}
\end{equation}

\noindent where $\mathrm{\mu_{0,c}}$ is the corrected central surface brightness value of the disk. Here, z =0.0485, as discussed in Section 3.1.

Finally, we use $\mathrm{\mu_{0,c}}$ of the disk, calculated for both g,r band images, to determine the equivalent B-band central surface brightness value using the equation given below \citep{Smithetal2002, PahwaSaha2018}. 

\begin{equation}
\mu_{0}(B) = \mu_{0,c}(g)+0.47(\mu_{0,c}(g)-\mu_{0,c}(r))+0.17  
\end{equation}

We present all these structural parameters (${\mu_{0,sersic}}$, $R_{e}$, n for S\'{e}rsic, and $\mu_{0,c}$, ${R_{D}}$ for disks corresponding to g, r, bands as well as $\mu_{0}(B)$) in Table \ref{tab:table1} corresponding to double-exponential modeling, and, in Table \ref{tab:table2} corresponding to single-exponential disk modeling. We note that the inner disk scale lengths in the double-disk model are slightly smaller than that of the single-disk model.

\begin{table*}
\centering
\caption{Structural parameters of the seven double-exponential disk galaxies, studied in this work, corresponding to S\'{e}rsic plus double-exponential disk modeling by GALFIT.}
\begin{tabular}{cccccccc}
\hline
\hline
Subhalo ID & \multicolumn{3}{c}{{Central component(Bulge/bar)}} & \multicolumn{2}{c}{Inner disk (ID)} & \multicolumn{2}{c}{Outer disk (OD)} \\
\hline
 & \multicolumn{3}{c}{S\'{e}rsic} & \multicolumn{2}{c}{Exponential} & \multicolumn{2}{c}{Exponential} \\
\hline
& $\mathrm{\mu_{0,sersic}}$ & $R_{e}$ & $n$  & $\mathrm{\mu_{0,c,ID}}$ & $R_{D,ID}$ & $\mathrm{\mu_{0,c,OD}}$ & $R_{D,OD}$  \\
&  (mag arcsec$^{-2}$) & (arcsec) &  & (mag arcsec$^{-2}$) & (arcsec)  & (mag arcsec$^{-2}$) & (arcsec) \\
\hline
(1)  & (2)   & (3)  & (4) & (5) & (6)  & (7) &(8) \\
\hline
&   &   & &  SDSS g-band (Optical B band)  &  &   &  \\
\hline
\hline
143882   & 15.97 & 1.7  & 2.1   & 20.2( 20.8) & 6.0 & 23.5(23.8) & 18.9 \\
220596   & 18.06  & 0.85 & 0.5  & 19.6(20.2) & 2.86 & 22.4(22.96) & 9.9  \\
242790   & 17.32 &  0.89 & 0.86  & 20.8(21.4) & 5.1 & 23.9(24.4) & 32.4 \\
421555   & 19.07 &  1.04 & 0.67  & 20.9(21.5) & 3.13 & 23.4(23.8) & 18.6 \\
450916   & 18.8   & 1.0  & 0.62 & 21.1(21.7) & 4.6 & 24.4(24.8) & 17.3 \\
452031   & 18.3 &  0.91 & 0.72  & 21.5( 22.0) & 6.2 & 23.8(24.6) & 12.75 \\
484448   & 18.7 &  0.87 & 0.50  & 20.7(21.2) & 2.7 & 23.4(24.1) & 10.2 \\
\hline
\hline
&   &   & & SDSS r-band   &   &    &  \\
\hline
\hline
143882   & 15.1 &  1.72  & 2.1   & 19.4  & 6.0  & 23.3 & 18.9\\
220596   & 17.1 &  0.86 & 0.5   & 18.8   & 2.86  & 21.6 & 9.9 \\
242790   & 16.4 &  0.89 & 0.86  & 19.94 & 5.11 & 23.3 & 32.4\\
421555   & 18.08 &  1.01 & 0.66  & 20.09 & 3.14 & 22.6 & 15.8 \\
450916   & 17.9 &  1.0  & 0.62  & 20.3  & 4.6  & 23.7 & 17.9 \\
452031   & 17.4 &  0.91 & 0.72  & 20.73 & 5.4 & 22.35 & 10.9 \\
484448   & 17.8 &  0.87 & 0.50  & 19.93 & 2.6 & 22.2 & 9.25 \\
\hline
\hline
\end{tabular}

Note: Column (1) denotes the Subhalo ID of the double-exponential disk galaxies. Cols.(2),(3) \& (4) represent the central surface brightness value, the effective radius, and the S\'{e}rsic index, respectively, of the best-fit S\'{e}rsic component. Cols.(5) \& (6) represent the central surface brightness value and the radial scale length, respectively, of the best-fit inner exponential disk. Cols.(7) \& (8) represent the same for the best-fit outer exponential disk. These values are given corresponding to the synthetic SDSS g \& r band images. The central surface brightness values of the disks calculated for the optical B band are shown as well.
\label{tab:table1}
\end{table*}

\begin{table*}
\centering
\caption{Same as Table \ref{tab:table1}, but for S\'{e}rsic plus single exponential disk model.}
\begin{tabular}{cccccc}
\hline
\hline
Subhalo ID & \multicolumn{3}{c}{{Central component(Bulge/bar)}} & \multicolumn{2}{c}{Disk} \\
\hline
& \multicolumn{3}{c}{S\'{e}rsic} & \multicolumn{2}{c}{Exponential} \\
\hline
& $\mathrm{\mu_{0,sersic}}$ & $R_{e}$ & $n$  & $\mu_{0,c}$ & $R_{D}$  \\
&  (mag arcsec$^{-2}$) & (arcsec) &  & (mag arcsec$^{-2}$) & (arcsec)\\
\hline
(1)  & (2)   & (3)  & (4) & (5) & (6) \\
\hline
&   &  &  & SDSS g-band(Optical B band)    &  \\
\hline 
\hline
143882   & 15.97 & 1.73 & 2.1  & 20.2(20.85) & 6.85 \\
220596   & 17.94  & 0.88 & 0.54  & 19.9(20.5) & 3.98 \\
242790   & 17.34 &  0.89 & 0.85  & 20.63(21.13) & 5.1\\
421555   & 19.1 &  1.19 & 0.67  & 21.7(22.4) & 7.16 \\
450916   & 18.76   & 1.01  & 0.62 & 21.19(21.86) & 5.5  \\
452031   & 18.32 &  0.91 & 0.72  & 21.44(22.04) & 7.12 \\
484448   & 18.7 &  0.87 & 0.50  & 20.84(21.4) & 3.57 \\
\hline
\hline
&   &  &  SDSS r-band   &  & \\
\hline
\hline
143882   & 15.26 & 1.67 & 1.99 & 19.3 & 5.95 \\
220596   & 17.01 & 0.89  & 0.54 & 19.03 & 3.84\\
242790   & 16.24 & 0.85 & 0.89 & 19.92 & 5.43 \\
421555   & 17.97 & 1.11 & 0.76 & 20.61 & 5.82 \\
450916   & 17.41 & 0.95 & 0.88 & 20.12 & 4.84 \\
452031   & 17.39 & 0.89 & 0.70 & 20.53 & 6.48 \\
484448   & 17.78 & 0.87 & 0.50 & 21.4 & 3.43 \\
\hline
\hline
\end{tabular}
\label{tab:table2}
\end{table*}

\begin{table*}
\centering
\caption{Redshift, Inner \& outer disk structural parameters of a sample of observed double-exponential disk galaxies taken from the references as mentioned in the footnote of the table. The central surface brightness ($\mathrm{\mu_{0}}$) and radial scale lengths of the inner ($\mathrm{R_{D,ID}}$) and outer ($\mathrm{R_{D,OD}}$) exponentials, typically measured in g, r, B bands are presented. We note that $\mathrm{\mu_{0}}$ for inner disk in Malin 1 is reported in HST/F814W band. In each case, the outer disk is concluded as a low surface brightness disk in the corresponding observed band.}
\begin{tabular}{ccccccc}
\hline
\hline
Galaxy ID & Redshift(z)  & \multicolumn{2}{c}{Inner disk} & \multicolumn{2}{c}{Outer disk}  \\
\hline
&  &  $\mathrm{\mu_{0,ID}}$ & $R_{D,ID}$ & $\mathrm{\mu_{0,OD}}$ & $R_{D,OD}$ \\
&  &  (mag arcsec$^{-2}$) & (kpc) & (mag arcsec$^{-2}$) & (kpc) & \\
\hline
Malin 1$^{a}$  & 0.0827 &  17.64       & 4.8  &  26.7  & 47.0  \\
UGC 1378$^{b}$ & 0.0098  &  19.88      & 4.51  &  21.54 & 12.6 \\
UGC 1382$^{c}$ & 0.0186  &  20.1      & 6.0 &   25.8 & 38.0 \\
NGC 2841$^{d}$ & 0.0021  &  $\sim$19.0 & 3.1  &  $\sim$25.0 & 13.0 \\
IZw 81$^{e}$   & 0.0518  &  20.31      & 2.04 &  23.32 & 7.35  \\
\hline
\hline
\end{tabular}
\begin{tablenotes}
\item [a] $^{a}$\citet{Sahaetal2021}, $^{b}$\citet{Saburovaetal2019}, $^{c}$\citet{Hagenetal2016}, $^{d}$\citet{Zhangetal2018},  $^{e}$\citet{Pandeyetal2022}
\end{tablenotes}
\label{tab:table3}
\end{table*}

A galaxy disk is identified as a low surface brightness disk if $\mu_{0}(B)$ is fainter than 22.5 $\mathrm{mag \ arcsec^{-2}}$ \citep{McGaugh1996,Rosenbaumetal2009}, which is a widely used criterion in literature. The values in Table \ref{tab:table1} show that the inner disks are HSB components, while the outer disks indeed satisfy the LSB disk determination criteria.  

Interestingly, the structural parameters of the HSB \& LSB disk components fall within the range of that of the observed double-disk galaxies, discussed in Section 1, as given in Table \ref{tab:table3}. 
In each case, the extended LSB disk has a radial scale length more than 10 kpc (except ID: 220596 and 484448, that have slightly smaller scale lengths). This shows that the simulated massive, double-exponential galaxies are morphologically good representative of the observed double-
disk galaxies. Therefore, we can consider the simulated spiral ones (143882, 242790, 421555, 452031) to be similar to the double-exponential spiral GLSBs as seen in observations. We also note that the simulated double-exponentials are not as extreme as Malin 1. Interestingly, we also have found 3 cases that appear to have lenticular morphology (220596, 450916, 484448). We note that the overall radial extents of the subhalo IDs: 220596 and 484448 are significantly smaller than that of a typical GLSB.

\subsection{Color profiles and star formation property}
\begin{figure}
\centering
\includegraphics[height=2.8in,width=3.7in]{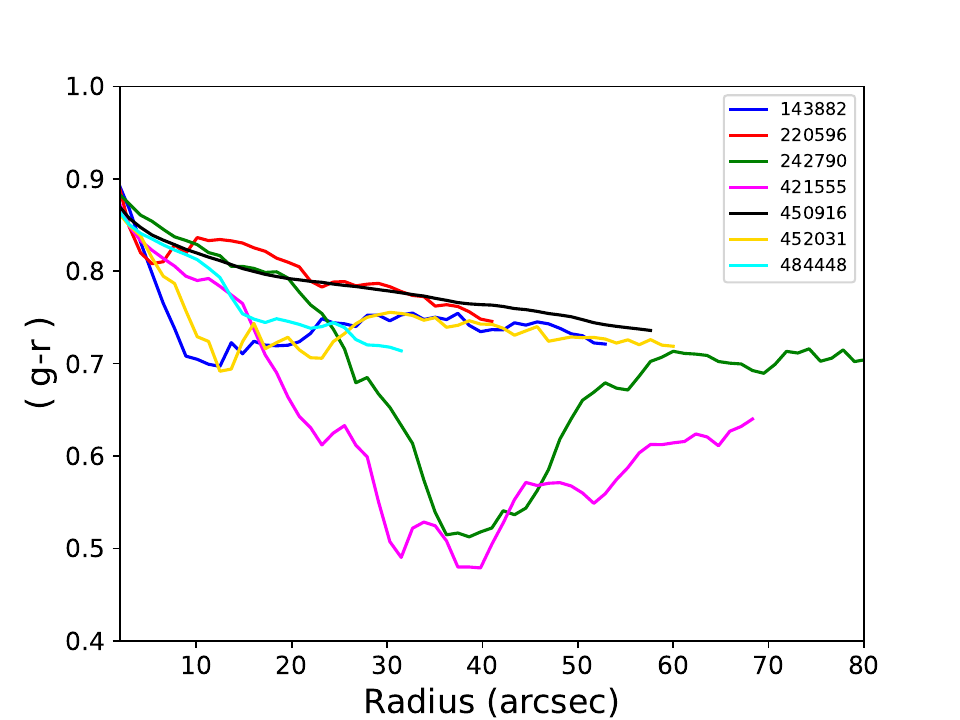}
\caption{Plot of g-r color radial profile of the double-disk galaxies obtained from their corresponding synthetic SDSS-g,r band images.}
\label{fig:fig5}
\end{figure}

\begin{figure}
\centering
\includegraphics[height=2.8in,width=3.4in]{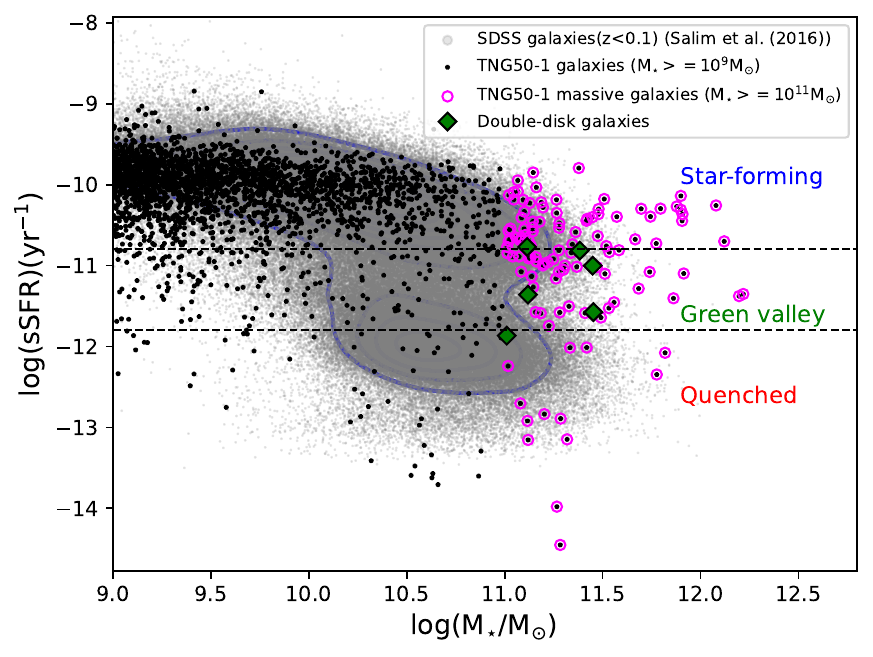}
\caption{Specific star formation rate (sSFR) vs. $\mathrm{M_{\star}}$: The background galaxies represented by grey dots are the observed local galaxies from SDSS observation. The sSFR vs $M_{\star}$ plane is divided into three regions: star-forming (log(sSFR)$>$=-10.8), green-valley (-10.8$>$log(sSFR)$>$-11.8) and quenched (log(sSFR)$<$=-11.8) \citep[see][]{BarwaySaha2020}. The black points represent all TNG50-1 galaxies with total $M_{\star}>10^{9}M_{\odot}$ with resolved SFR. The 132 massive galaxies used in our work are marked by magenta circles. The double-disk galaxies are marked by the green diamonds among them. The subhalo ID: 421555 lies at the edge of the star-forming region, ID: 484448 is in the quenched region. The rest of the double-disk galaxies (143882, 220596, 242790, 452031) are in the green-valley region except ID: 450916, which has an unresolved SFR value and is, therefore, assigned SFR=0 in the data catalog. It is noted that all the massive TNG50-1 galaxies are selected with a mass cut-off of $\mathrm{log(M_{\star}/M_{\odot})>=11.0}$. All the double-disk galaxies lie above this mass cut-off in the figure.}\label{fig:fig6}
\end{figure} 

The radial color profiles of disk galaxies are often discussed in the literature to understand the formation and evolution of the galaxies. Because the profiles reflect the distribution of stellar populations in age, metallicity as well as the star-formation history of the galaxies. They are also often related to the slope observed in the radial surface brightness profiles \citep{Bakosetal2008,Herrmannetal2016}. In literature, typical local spirals are observed to have a redder bulge and a bluer disk component \citep{Morsellietal2017,Barsantietal2021}.

We obtain the radial profile of the (g-r) optical color for each simulated double-exponential galaxy (Figure \ref{fig:fig5}) using the average flux per pixel calculated within the concentric circular annulus placed over the corresponding SDSS-(g-r) color map. We note that the color profiles usually decrease from a maximum at the center to a minimum and then rise again and attain a flat part. Overall, the central part of each galaxy appears to be the reddest, with (g-r)$\ge$ 0.8 in all cases studied here - indicating a redder bulge/bar component associated with a bluer disk. The striking feature of some of these double-disk galaxies appears to be their color minima.  We do not find any apparent correlation between the radii corresponding to the color minima and the radii from which LSB disks start to dominate the surface brightness profile.

In the following, we explore the star formation property of the double-disk galaxies. The specific star formation rate, i.e., the ratio of the rate of current star formation to the stellar mass, shows a bi-modality in the sSFR vs. $\mathrm{M_{star}}$ plane, also manifested in their color-magnitude relation for the observed local galaxies \citep{Kauffmann2003,Baldryetal2006,DekelBirnboim2006,Ilbertetal2015}. 
On the sSFR vs. $\mathrm{M_{star}}$ plane, the main sequence star-forming galaxies lie above log(sSFR) $>=-10.8 yr^{-1}$, the quenched galaxies with very little or no star formation lie below log(sSFR)$=<-11.8 yr^{-1}$. The galaxies lying between these two regions are termed the green-valley galaxies \citep{Salim2014,Schawinskietal2014,Baitetal2017,BarwaySaha2020,Abrahametal2024,Goddardetal2024,Chandler2024}. \citet{Salim2014} show that these green-valley galaxies also lie well within the green-valley region defined on the NUV-r vs. Magnitude-r plane.

We show the simulated double-disk galaxies, the sample of 132 massive galaxies, and all the simulated galaxies with $\mathrm{M_{star}}>10^{9}M_{\odot}$ from TNG50-1 on the sSFR vs. $\mathrm{M_{star}}$ plane in the backdrop of observed local galaxies in Figure \ref{fig:fig6}. The data for the observed galaxies from SDSS observation are taken from the catalog by \citet{Salimetal2016}. The stellar mass, and SFR of those galaxies calculated as a time-averaged quantity over the last 100 Myr, are obtained from SED modeling (along with extinction correction) in the catalog. The density contours plotted on this background sample clearly show the bi-modality in sSFR. We note that the error corresponding to the SFR of the observed SDSS galaxies ranges between 0.65 and 0.80 dex for the passive galaxies, and is typically below 0.1 dex for the star-forming galaxies. Errors on the stellar mass range from 0.03 dex for the passive galaxies to 0.10 dex for the most active ones \citep{Salimetal2016}. The median error in log(stellar mass) is 0.036 $M_{\odot}$, in log(SFR) is 0.175 $M_{\odot}yr^{-1}$.

The SFR data of all the simulated galaxies are taken from the supplementary data catalog available on the TNG website \citep{Donnarietal2019, Pillepichetal2019}. 
The mass of each gravitationally bound star particle of the galaxy and the formation time of each star particle are known from the simulation data. Using those values, the SFR is measured as a time-averaged quantity over different time periods as well as over different apertures. We consider the subhalo SFR calculated using all of the gravitationally bound stars of the subhalo, time-averaged across the last 100 Myr (similar to the observed galaxies) from the above catalog. Among the sample of the double-disk galaxies, we find one (ID:421555) to be just at the edge of the star-forming region, one (Subhalo ID: 484448) to be in the quenched region, and the rest of the four galaxies (Subhalo IDs: 143882, 220596, 242790, 452031) to lie in the green-valley region. The subhalo ID: 450916 does not have a resolved SFR value due to the finite numerical resolution of the simulation. It has been artificially assigned a zero value for SFR in the catalog \citep{Donnarietal2019}, and therefore it does not appear in the plot here. We also note that the median of the stellar mass distribution of the simulated massive disk galaxies, lying in the green-valley region, turns out to be $\mathrm{log(M/M_{\odot}})=11.3$, and the maximum turns out to be $\mathrm{log(M/M_{\odot}})=12.2$. 
Although, our selection was not based on color, an analysis based on the SDSS observations of z=0 galaxies suggests that there exists a critical mass limit of $\mathrm{log(M/M_{\odot}})=11.7$ beyond which no spiral galaxies can exist \citep{Martinetal2018, Quilleyetal2022}. Further, \citet{Quilleyetal2022} showed that their green plain has the highest mass disk galaxy at $\mathrm{log(M/M_{\odot}})=11.7$ beyond which the ellipticals dominate. The fact that we see massive disk galaxies with $\mathrm{log(M/M_{\odot}})\ge 11.7$ both in the star-forming and green-valley region in TNG50-1 at z=0, indicates that they might have managed to escape a major merger event in the recent past. This requires a further follow up. 
We note that the sample of TNG50-1 galaxies with $M_{\star}>10^{9}\mathrm{M_{\odot}}$ do not show a strong bi-modality in sSFR (for a detailed discussion, see \citet{Donnarietal2019}). 

\subsection{Tully-Fisher relation}
\begin{figure}
\centering
\includegraphics[height=3.2in,width=3.8in]{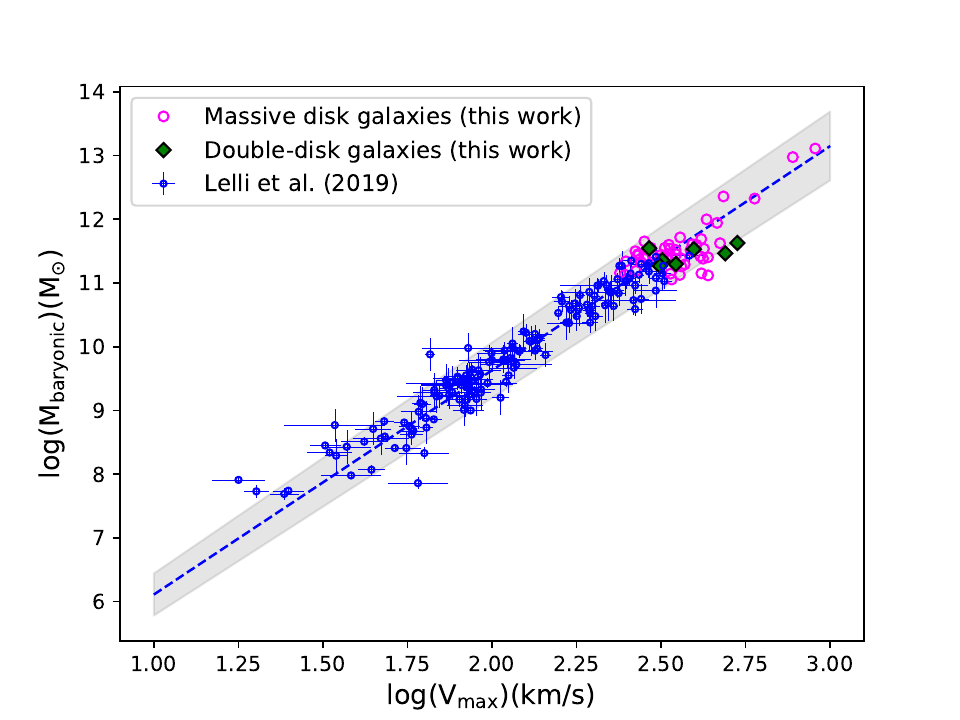}
\caption{Baryonic Tully-Fisher relation: The magenta circles represent the sample of 57 massive simulated disk galaxies used in our work. We mark the double-disk galaxies among them with green diamonds. The blue circles represent observed disk galaxies, studied by \citet{Lellietal2019}. This sample follows the BTFR (blue dashed line) with a slope of 3.52$\pm$ 0.07 \& an intercept of 2.59$\pm$0.15, as reported in their paper. The grey shaded region represents 1.5-$\sigma$ region of the above observed BTFR.}
\label{fig:fig7}
\end{figure}
The Tully-Fisher relation \citep{TullyFisher1977,BelldeJong2001,Courteauetal2003,denHeijeretal2015} is the correlation between the mass of a disk galaxy and its rotation velocity. This has been shown to be one of the most fundamental galaxy scaling relations for disk galaxies, in observations, so far. This can be studied in terms of stellar mass, i.e., galaxy luminosity vs. rotation velocity \citep{TullyFisher1977} as well as the total baryonic mass (stars+gas) vs. rotation velocity. The baryonic Tully-Fisher relation provides a much stronger correlation than the stellar mass Tully-Fisher relation, with much reduced scatter \citep{McGaughetal2000,Gurovichetal2004, McGaugh2005,Trachternachetal2009}. The linear slope of the BTFR is usually observed to lie in the range of 3-4. In this work, we explore the BTFR of the massive disk (57) galaxies in our sample, including the double-disk cases, in Figure \ref{fig:fig7}. As the observational sample, we consider the late-type disk galaxies studied in \citet{Lellietal2019}, based on the SPARC dataset \citep{Lelli_sparc_2016}. Using the maximum rotation velocity ($\mathrm{V_{max}}$), the slope of the BTFR comes out to be 3.52 $\pm$ 0.07 in their paper. We show the 1.5-$\sigma$ region (grey shaded region) of the observed BTFR, calculated using the corresponding values reported in their paper, in Figure \ref{fig:fig7}. We use the gravitationally bound total baryonic mass (stars+gas), and $V_{max}$, i.e., the maximum of the spherically averaged rotation curve of the simulated galaxies available in the TNG SUBFIND data catalog, and show these on the same figure. Interestingly, all the double-disk galaxies are found to lie within the 1.5-$\sigma$ region of the observed BTFR, except two (Subhalo IDs: 143882, 220596) which lie just at the edge of the region. 

\section{Conclusion}
We have studied massive disk galaxies from the highest resolution run of IllustrisTNG50 simulation in order to find galaxies with a double-exponential disk morphology where the outer disk is a low surface brightness component surrounding the central high surface brightness inner disk component. We used the idealized, synthetic SDSS g, r band images of the galaxies , available in the TNG supplementary data catalog, to do detailed 2D structural decomposition. We summarize the main results from our paper as follows.
\begin{itemize}
\item
We find seven disk galaxies in TNG50 that are best described by S\'{e}rsic plus two exponential (HSB plus extended LSB) disks. This sample is $\sim$ 12\% of the 57 disk galaxies from TNG50 chosen for GALFIT modeling. The single-exponential disk model clearly leaves systematic, significant flux in the outskirts of these seven galaxies. The addition of an outer exponential disk is found to model that flux and improve the goodness of the fit. This is similar to the observed massive GLSBs that show such hybrid morphology.
 
\item 
The central surface brightness values of the inner and outer exponential disks lie in the range of 20.2- 22.0 $\mathrm{mag ~arcsec^{-2}}$ \& 23-24.8 $\mathrm{mag~arcsec^{-2}}$, respectively, in optical B-band. The radial scale lengths of the disks lie in the range of 2.6-6.1 kpc \& 9.7-31.7 kpc, respectively, corresponding to the SDSS-g band. These values are in agreement with those in observed double-exponential disk galaxies.

\item
Using TNG supplementary data catalog, we studied the time-averaged (over the last 100 Myr) specific star formation rate of the double-disk galaxies. One of the double-disk galaxies lies at the edge of the star-forming region, one lies in the quenched region, and the rest of the sample lie in the green-valley region compared to the observed local ($z<0.1$) SDSS galaxies. We found one double-disk galaxy in our sample to have an unresolved SFR value.

\item
We measure radial optical color profiles of the galaxies, which usually show a global minimum at the center. There is no apparent correlation between the radii corresponding to the color minima and the radii from which the LSB disks start to dominate the surface brightness profile. 

\item 
Interestingly, the double-disk galaxies, along with the massive disk galaxies, are found to lie within the 1.5$\sigma$ region of the observed Baryonic Tully-Fisher relation.
\end{itemize}

In summary, this work using cosmological simulation data presents a detailed structural analysis of massive double-disk galaxies, bringing out a couple of interesting points that help in improving our understanding of the morphology of the massive disk galaxies so far. Our work shows that massive disk galaxies can have an outer, extended LSB envelope around their central HSB core. A detailed light profile modeling up to the fainter outskirt can reveal that. Similarly, it also shows that the GLSBs with extended LSB disk structure could actually consist of a HSB plus LSB disk structure.


The results from our work along with the observations in the literature, invoke the following questions: How common are such galaxies in the local universe? Do massive disk galaxies in the nearby universe grow in size and mass by developing an outer, extended LSB envelope? How does the faint, extended disk survive? A generic understanding of these would be possible by analyzing a statistically large sample.

Our results obtained from simulations could serve as the motivation to look for such galaxies in deep photometric observations such as from HSC-SSP, 
Dragonfly observation 
as well as from the upcoming data from LSST 
as mentioned in Section 1. 
We note that HSC-SSP can go up to  $\sim$ 28 mag in r-band. Dragonfly telescope can observe up to the 1-D surface brightness level of $\sim$ 32 $\mathrm{mag ~arcsec^{-2}}$ \citep{vanDokkumetal2014,Zhangetal2018}. On the other hand, LSST, with its 10-year survey, promises to provide the detection of billions of galaxies and very faint structures. Interestingly, the surface brightness profiles of the double-disk galaxies in our work are found to reach typically $\sim$ 28 $\mathrm{mag ~arcsec^{-2}}$. Thus, with the sensitivity of the above mentioned telescopes, it will be possible to perform targeted observation of such galaxies as well as build a statistically large sample. This will facilitate a detailed comparison of morphology, the extended LSB disk structure of such galaxies with those studied from simulations.

As a future work, we intend to study these simulated double-disk galaxies across redshift to detect the formation and evolution of the outer LSB envelope by doing a similar morphological analysis. 
We have also explored the massive dark haloes of the double-disk galaxies in the IAU proceedings corresponding to the IAU general assembly, 2024.

\section*{Acknowledgments}
We thank the anonymous referee for very useful comments. SS thanks Divya Pandey, Suraj Dhiwar, Anshuman Borgohain for useful discussions. The research in this article has been carried out using the IllustrisTNG simulation data publicly available on the IllustrisTNG website, the data from the publicly available SPARC (Spitzer Photometry \& Accurate Rotation Curves) database \citep{Lelli_sparc_2016}, and from SDSS. Software: GALFIT, IRAF, Astropy, DS9.

\bibliography{ms}{}
\bibliographystyle{aasjournal}

\end{document}